\documentclass[12pt,preprint]{aastex}
\usepackage{graphicx}
\usepackage{longtable}
\usepackage{url}
\usepackage{lipsum}

\shorttitle{Low-redshift excess of Swift GRB rate} \shortauthors{Yu
et al.}

\slugcomment{}

\begin{document}
%\title{The luminosity function and formation rate of \emph{Swift} gamma-ray bursts}

\title{An unexpectedly low-redshift excess of \emph{Swift} gamma-ray burst rate}

\author{H. Yu\altaffilmark{1}, F. Y. Wang\altaffilmark{1,2,3}{*}, Z. G. Dai\altaffilmark{1,2} and K. S. Cheng\altaffilmark{3}}
\affil{
$^1$ School of Astronomy and Space Science, Nanjing University, Nanjing 210093, China\\
$^2$ Key Laboratory of Modern Astronomy and Astrophysics (Nanjing University), Ministry of Education, Nanjing 210093, China\\
$^3$ Department of Physics, University of Hong Kong, Pokfulam Road,
Hong Kong, China\\ } \email{{*}fayinwang@nju.edu.cn}

\begin{abstract}
Gamma-ray bursts (GRBs) are the most violent explosions in the
Universe and can be used to explore the properties of high-redshift
universe. It is believed that the long GRBs are associated with the
deaths of massive stars. So it is possible to use GRBs to
investigate the star formation rate (SFR). In this paper, we use
Lynden-Bell's $c^-$ method to study the luminosity function and rate
of \emph{Swift} long GRBs without any assumptions. We find that the
luminosity of GRBs evolves with redshift as $L(z)\propto
g(z)=(1+z)^k$ with $k=2.43_{-0.38}^{+0.41}$. After correcting the
redshift evolution through $L_0(z)=L(z)/g(z)$, the luminosity
function can be expressed as $\psi(L_0)\propto L_0^{-0.14\pm0.02}$
for dim GRBs and $\psi(L_0)\propto L_0^{-0.70\pm0.03}$ for bright
GRBs, with the break point $L_{0}^{b}=1.43\times10^{51}~{\rm
erg~s^{-1}}$. We also find that the formation rate of GRBs is almost
constant at $z<1.0$ for the first time, which is remarkably
different from the SFR. At $z>1.0$, the formation rate of GRB is
consistent with the SFR. Our results are dramatically different from
previous studies. Some possible reasons for this low-redshift excess
are discussed. We also test the robustness of our results with Monte
Carlo simulations. The distributions of mock data (i.e.,
luminosity-redshift distribution, luminosity function, cumulative
distribution and $\log N-\log S$ distribution) are in good agreement
with the observations. Besides, we also find that there are
remarkable difference between the mock data and the observations if
long GRB are unbiased tracers of SFR at $z<1.0$.
\end{abstract}


\section{Introduction}
Gamma-ray bursts (GRBs) are a kind of the most violent explosions in
the Universe, which radiate huge energy in gamma-ray in a short time
\citep[for reviews, see][]{Meszaros06,Zhang07,Gehrels09}. They are
so bright and can be detected at much higher redshifts than
supernovae (SNe). Hitherto, the farthest GRB with spectroscopic
redshift is GRB 090423 at $z\approx8.2$
\citep{Tanvir09,Salvaterra09}. So GRBs may be useful tools to probe
the early universe \citep{Bromm12,Wang14}, including dark energy
\citep{Dai04,Schaefer07,Wang11}, star formation rate
\citep{Totani97,Wang13}, reionization \citep{Totani06,McQuinn08},
and metal enrichment \citep{Wang12,CastroTirado14}.

Theoretically, the progenitors of long GRBs with duration time
$T_{90}>2$ s are thought to be collapse of massive stars
\citep{Woosley93}. Observations also show that some long GRBs are
associated with the deaths of massive stars which will give core
collapse supernovae \citep{Stanek2003,Hjorth2003}. So GRBs can be
used to investigate the star formation rate (SFR) at high redshifts
\citep{Totani97,Wijers1998,LambReichart2000,PorcianiMadau2001,BrommLoeb2002,Lin04,Kistler2009,Wang09,Wanderman2010,Butler2010,Elliott12,Elliott14}.
In order to measure SFR by using GRBs, the relation between the rate of
GRBs and SFR should be known. Some studies found that the GRB rate
is consistent with SFR at about $z<4.0$, but has an excess at high
redshift comparing with that expected from SFR
\citep{Le07,Kistler2009}. Some models have been proposed to explain
the discrepancy between SFR and GRB rate, such as the cosmic
metallicity evolution \citep{Li08,Qin10}, superconducting cosmic
strings \citep{Cheng10}, evolving initial mass function (IMF) of
stars \citep{WangD11}, evolution of the GRB luminosity function
break \citep{Virgili11}. From the redshift distribution of GRBs and
the metallicity of GRB host galaxies, \citet{WangD14} showed that
the discrepancy between the GRB rate and SFR can be reconciled by
considering that GRBs occur in low-metallicity galaxies.

In previous literatures, the $\log N-\log P$ distribution has been
used to study the luminosity function and formation rate of GRBs
\citep{FenimoreRamireRuiz2000,Firmani2004,Guetta05,Guetta07,Salvaterra07,Salvaterra09b,Cao11}.
But the $\log N-\log P$ distribution is the production of the
intrinsic luminosity function convolving with the formation rate of
GRBs, so the luminosity function and formation rate of GRBs are
degenerated \citep{Firmani2004,Guetta07,Howell14}.

Moreover, there are several selection effects on the observed
redshift distribution of GRBs \citep{Coward07}, and so the rate of
GRBs. The most important one is the observational limit of
satellite. The \emph{Swift} satellite has a flux limit, which means
that it can not detect GRB dimmer than the flux limit. So the
observed data is truncated, it will be difficult to get the
intrinsic distribution of GRB before the selection effect is
corrected. Lynden Bell (1971) applied a novel method to study the
luminosity function and density evolution from flux-limit quasar
sample, which is called Lynden-Bell's $c^{-}$ method
\citep{Lynden-Bell1971}. After that, this method has been widely
used in some other objects, such as galaxies
\citep{Kirshner1978,Merighi1986,Peterson1986,Loh1986}, long GRBs
\citep{Yonetoku2004,Lloyd-Ronning2002,Kocevski06,Wu2012}, and short
GRBs \citep{Yonetoku14}. The basis of Lynden-Bell's $c^{-}$ method
is that the distributions of luminosity $L$ and redshift $z$ are
independent \citep{Lynden-Bell1971,EfronPetrosian1992}. So before
applying this method, we need to test the independence of $L$ and
$z$ with a nonparametric test method provided by
\citet{EfronPetrosian1992}. Lynden-Bell's $c^{-}$ method is a
powerful method to estimate the luminosity function and formation
rate of objects with truncated sample. For example,
\citet{Yonetoku2004} used this method to derive the luminosity
function and the formation rate of GRB from 689 \emph{BATSE} GRBs
with pseudo redshifts. They found that the GRB formation rate
increases quickly at $0 < z < 1$, and remains approximately constant
up to $z \sim 10$, which is consistent with observed SFR at $z<4.0$.
\citet{Kocevski06} found that the GRB comoving rate density rises
steeply at $z<1.0$, followed by flattening and declines at about
$z>3.0$. \citet{Wu2012} also studied the formation rate of 95
\emph{Swift} GRBs using Lynden-Bell's $c^{-}$ method, and found that
the GRB formation rate increases quickly at $0 < z < 1.0$, and
remains approximately constant at $1.0 < z < 4.0$, but finally
decreases at $z > 4.0$, which is well consistent with the SFR
\citep{Hopkins2006,Kistler2009,Yuksel2008,Wang09}.

In this paper, we study luminosity function and formation rate of
latest \emph{Swift} GRBs by using Lynden-Bell's $c^{-}$ method. The
Lynden-Bell's $c^{-}$ method can break the degeneracy between
luminosity function and GRB formation rate. This paper is organized
as follows. We introduce the data from \emph{Swift} satellite, and
make the K-correction in the next section. The introductions to
Lynden-Bell's $c^{-}$ method and the nonparametric $\tau$
statistical method are given in section 3. In section 4, we derive
the luminosity function and formation rate of GRBs. Monte Carlo
simulation is used to test our results in section 5. Finally,
section 6 presents conclusions and discussions. Throughout the
paper, we assume a flat $\Lambda$CDM universe with
${\Omega}_{m}=0.27$ and $H_{0}=70~{\rm km~s^{-1}Mpc^{-1}}$.

\section{GRB sample}\label{sec:data}
\emph{Swift} is a multi-wavelength satellite to observe GRBs. It has
instruments designed to analyze the bursts, X-ray and UV/optical
afterglows, which can locate the positions of GRBs. We collect 127
long GRBs with well measured spectral parameters given by
\emph{Fermi-GBM} and \emph{Konus-wind}. These GRBs have redshift
data observed by \emph{Swift}. In Table 1, we list the GRB sample,
including name (Column 1), redshift (Column 2), low-energy power-law
index $\alpha$ (Column 3), high-energy power-law index $\beta$
(Column 4), peak energy of the $\nu F_\nu$ spectrum in observer's
frame (Column 5), peak flux in a certain energy range (Column 6),
energy range (Column 7), bolometric luminosity (Column 8) and
references (Column 9) of GRBs.

Two spectral models are used to fit the spectra of GRBs, including a
power law with an exponential cutoff model (PLEXP) and the Band
model \citep{Band93}. The functional forms are as follows,

\begin{equation}\label{powerLawCutoff}
f(E) = A(\frac{E}{100~{\rm
keV}})^{\alpha}\exp^{-\frac{(2+\alpha)E}{E_{\rm p}}},
\end{equation}

\begin{equation}\label{BandModel}
f(E) = \left\{
\begin{array}{ll}
        A(\frac{E}{100~{\rm keV}})^{\alpha}
        \exp{- \frac{(2+\alpha)E}{E_{\rm p}}} \ \
        & E<\frac{(\alpha-\beta)E_{\rm p}}{2+\alpha}, \\
        A(\frac{E}{100~{\rm keV}})^{\beta}\exp^{\beta-\alpha}(
        \frac{(\alpha - \beta) E_{\rm p}}{(2+\alpha)100~{\rm keV}})^{\alpha - \beta} \ \
        & E\geq \frac{(\alpha-\beta)E_{\rm p}}{2+\alpha}, \\
\end{array}
\right.
\end{equation}
which represent a power law with an exponential cutoff model and the
Band model respectively.

Because the peak fluxes are observed in a lager range of redshifts
which correspond to different range of energy bands in the rest
frame of GRBs. The K-correction is required to get the bolometric
luminosity of GRBs \citep{Bloom01}. The bolometric luminosity of GRB
is
\begin{equation}
L = 4{\pi}d_{L}^{2}(z)FK,
\end{equation}
where
\begin{equation}
d_{L}(z)=\frac{c}{H_0}\int_0^z
\frac{dz}{\sqrt{1-\Omega_{m}+\Omega_{m}(1+z)^3}}
\end{equation}
is the luminosity distance at redshift $z$, $F$ is the peak flux
observed between between a certain energy range $(E_{min},E_{max})$,
and the $K$ is the factor of K-correction. If the flux $F$ is in
units of $\rm erg~cm^{-2}~s^{-1}$, the parameter $K$ is defined as
\begin{equation}\label{kCorrection1}
K = \frac{{\int}^{10^4{\rm keV}/(1+z)}_{1{\rm keV}/(1+z)}
Ef(E)dE}{{\int}^{E_{max}}_{E_{min}} Ef(E)dE}.
\end{equation}
If the flux $F$ is in units of $\rm photons~cm^{-2}~s^{-1}$, the
parameter $K$ is defined as
\begin{equation}\label{kCorrection2}
K = \frac{{\int}^{10^4{\rm keV}/(1+z)}_{1{\rm keV}/(1+z)}
Ef(E)dE}{{\int}^{E_{max}}_{E_{min}} f(E)dE},
\end{equation}
where $f(E)$ is spectral model of GRB. Then we can obtain the
bolometric luminosity $L$ of each GRB. In Figure
\ref{LyndenBellfig}, the blue dots show the bolometric luminosity of
GRBs, and the line is the observational limit of \emph{Swift}. The
limit is chosen as a minimum flux $F_{\rm min}=2.0\times10^{-8}~{\rm
erg~s^{-1}cm^{-2}}$, which is consistent with that of \citet{Li08}.
So the limit luminosity at redshift $z$ is given as $L_{\rm limit} =
4{\pi}d_{\rm L}^{2}(z)F_{\rm min}$.

\section{Lynden-Bell's $c^{-}$ method and nonparametric test method}\label{sec:method}
Lynden-Bell's $c^{-}$ method is an efficient method to determinate
the distribution of luminosity and redshift of objects with
truncated data sample, including quasars
\citep{Lynden-Bell1971,EfronPetrosian1992,Petrosian1993,MaloneyPetrosian1999},
galaxies \citep{Kirshner1978,Merighi1986,Peterson1986,Loh1986} and
GRBs \citep{Lloyd-Ronning2002,Yonetoku2004,Wu2012,Yonetoku14}.
\citet{Lynden-Bell1971} used this method to derive the luminosity
function and density evolution from quasars with observational
selection for the first time. This method can break the degeneracy
between luminosity function and formation rate. It is better to
extract the luminosity evolution from the form of the luminosity
function. If the parameters $L$ and $z$ in the distribution of
luminosity and redshift $\Psi(L,z)$ are independent, we can rewrite
$\Psi(L,z)$ as $\Psi(L,z)=\psi(L)\phi(z)$
\citep{EfronPetrosian1992}, where $\psi(L)$, the fraction of GRB
brighter than $L$, is the cumulative luminosity function, and
$\phi(z)$ is the redshift cumulative distribution. Unfortunately,
the luminosity and the redshift of GRBs are not independent
\citep{Yonetoku2004,Kocevski06,Wu2012}, so we should write the
$\Psi(L,z)$ as $\Psi(L,z)=\psi_z(L)\phi(z)$ instead of
$\Psi(L,z)=\psi(L)\phi(z)$, where $\psi_z(L)$ is the luminosity
function of GRB at redshift $z$. If we remove the effect of the
luminosity evolution $g(z)$, i.e. a transformation $L_{0}=L/g(z)$,
the transformed luminosity $L_0$ are independent of redshift. As a
result, we can obtain $\Psi(L_0,z)=\phi(z)\psi(L_0)$. Using the
relation $L = L_0g(z)$, we can write the $\Psi(L,z)$ as
$\Psi(L,z)=\psi_z(L)\phi(z)=\psi(L_0)\phi(z)$. Then the luminosity
function of GRB at redshift $z$ is $\psi_z(L)=\psi(L/g(z))$.

\begin{figure}
  % Requires \usepackage{graphicx}
  \centering
  \includegraphics[width=0.8\textwidth]{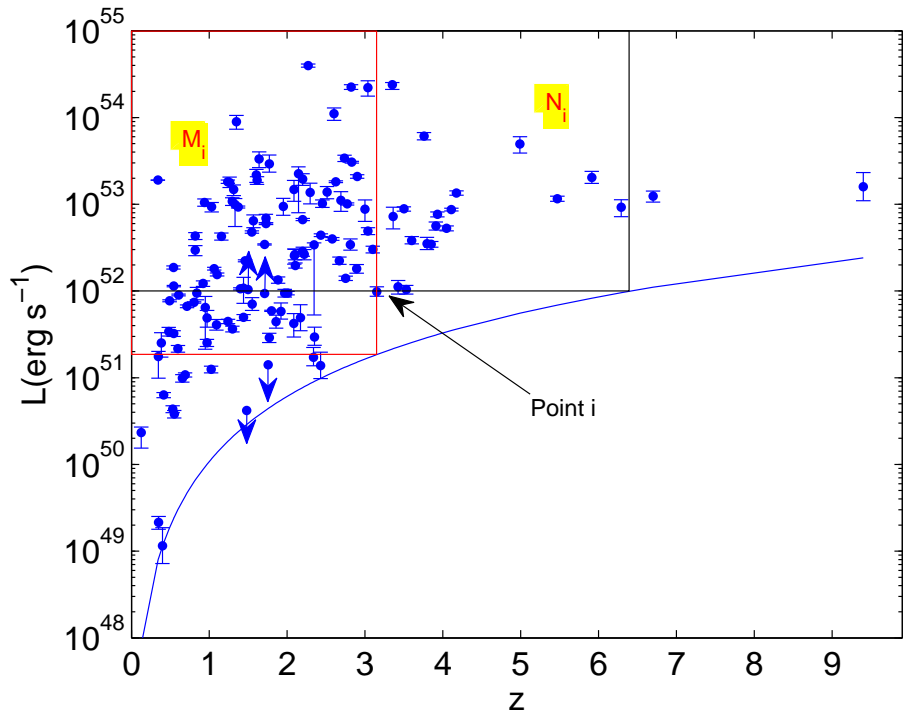}\\
  \caption{The luminosity distribution of 127 GRBs after K-correction. The blue dots
  represent GRBs and the blue line represents observational limit of \emph{Swift}. We take the flux limit as
  $2.0\times 10^{-8}~ \rm erg~cm^{-2}~s^{-1}$. $M_i$ and $N_i$ are also shown. The error bars are $1\sigma$ errors.}\label{LyndenBellfig}
\end{figure}

To get the form of luminosity evolution $g(z)$, we introduce a
nonparametric test method proposed by \citet{EfronPetrosian1992},
which has been widely used in previous literature
\citep{Petrosian1993,MaloneyPetrosian1999,Lloyd-Ronning2002,Yonetoku2004,Wu2012,Yonetoku14}.
For the $i$th data in the ($L,z$) data set, we can define $J_i$ as
\citep{EfronPetrosian1992}
\begin{equation}\label{Ji}
J_i = \{j|L_j \geq L_i , z_j\leq z_{i}^{max}\},
\end{equation}
where $L_i$ is the $i$th GBR luminosity and $z_{i}^{\rm max}$ is the
maximum redshift at which a GRB with luminosity $L_i$ can be
observed. This region is shown in Figure \ref{LyndenBellfig} as
black line rectangle. The number of GRBs contained in this region is
$n_i$. The number $N_i = n_i - 1$, which means taking the $i$th GRB out, is the same as $c^{-}$ in
\citet{Lynden-Bell1971}. Similarly, $J^{\prime}_i$ is defined as
\begin{equation}\label{JI}
J^{\prime}_i = \{j|L_j \geq L_{i}^{lim} , z_j < z_{i}\},
\end{equation}
where $L_{i}^{\rm lim}$ is the minimum observable luminosity at that
redshift $z_i$. This region is shown as red rectangle in Figure
\ref{LyndenBellfig}. The number of events contained within this
region is $M_i$.

We first consider the $n_i$ GRBs in black rectangle in Figure
\ref{LyndenBellfig}. The number of events that have redshift $z$
less than or equal to $z_i$ is defined as $R_i$. If $L$ and $z$ are
independent, $R_i$ is uniformly distributed between 1 and $n_i$
\citep{EfronPetrosian1992}. The test statistic $\tau$ is
\citep{EfronPetrosian1992}
\begin{equation}\label{statisticData}
\tau \equiv \sum_{i} \frac{(R_i - E_i)}{\sqrt{V_i}},
\end{equation}
where $E_i = \frac{1+n_i}{2}$, $V_i = \frac{n_i^2 - 1}{12}$ are the
expected mean and the variance of $R_i$ respectively. If the $R_i$
is exactly uniformly distributed between 1 to $n_i$, the samples of
$R_i \leq E_i$ and $R_i \geq E_i$ should be nearly equal, and the
test statistic $\tau$ will nearly be 0. If we choose a from of
$g(z)$ that makes test statistic $\tau=0$, the effect of the
luminosity evolution can be removed by transformation of
$L_{0}=L/g(z)$.

\begin{figure}
\centering
  \includegraphics[width=0.8\textwidth]{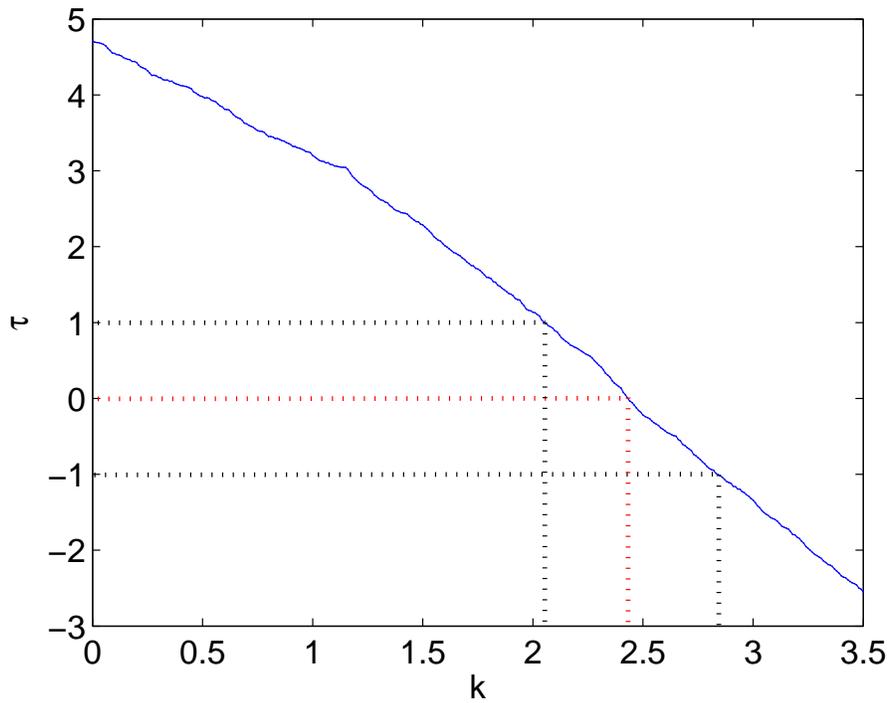}\\
  \caption{The value of test statistic $\tau$ as a function of $k$. The red dotted line represents the best
  fit for $\tau=0$, and the black dotted lines are $1\sigma$ errors. The value of $k$ is $k=2.43_{-0.38}^{+0.41}$
  at $1\sigma$ confidence level. It also shows that $\tau=4.7$ when $k=0$, which means $k=0$ is excluded at $4.7\sigma$ confidence level.}\label{kVakuefig}
\end{figure}

The functional form of $g(z)=(1+z)^k$ has been used in previous
papers
\citep{Lloyd-Ronning2002,Yonetoku2004,Wu2012,Lloyd-Ronning2002,Kocevski06,Yonetoku14}.
We also use this form in this paper. Then we test the independence
between $L_{0}=L/g(z)$ and $z$ by changing the value of $k$ until
the test statistic $\tau$ is zero. Figure \ref{kVakuefig} shows the
value of test statistic $\tau$ as a function of $k$. From this
figure, we find the best fit is $k=2.43_{-0.38}^{+0.41}$ at
$1\sigma$ confidence level. So we take the luminosity evolution form
$g(z)$ as $g(z)=(1+z)^{2.43}$. A hypothesis of no luminosity
evolution $k=0$, is rejected at about $4.7\sigma$ confidence level.
This value is similar with \citet{Yonetoku2004}, whose $k$-value is
$k=2.60_{-0.20}^{+0.15}$ and $k=0$ is rejected with about
$8.0\sigma$ significance from pseudo-redshift GRBs. \citet{Wu2012}
also found that the value of $k$ is $k=2.30_{-0.51}^{+0.56}$.

After converting the observed luminosity to non-evolving luminosity
$L_{0}=L/(1+z)^{2.43}$, we can derive the local cumulative
luminosity function $\psi (L_0)$ with nonparametric method from the
following equation \citep{Lynden-Bell1971,EfronPetrosian1992},
\begin{equation}\label{LuminosityFunction}
\psi(L_{0i}) = \prod\limits_{j<i}(1+\frac{1}{N_j}),
\end{equation}
where $j<i$ means that GRB has luminosity $L_{0j}$ larger than
$L_{0i}$. The cumulative number distribution $\phi(z)$ can be
obtained from
\begin{equation}\label{NumberFunction}
\phi(z_i) = \prod\limits_{j<i}(1+\frac{1}{M_j}),
\end{equation}
where $j<i$ means that GRB has redshift $z_j$ less than $z_i$. The
comoving differential form of $\phi(z)$, which represents the cosmic
formation rate of GRBs $\rho(z)$, is more interesting. The formation
rate of GRBs can be written as,
\begin{equation}\label{formationrate}
\rho(z) = \frac{d\phi(z)}{dz}(1+z)(\frac{dV(z)}{dz})^{-1},
\end{equation}
where $(1 + z)$ results from the cosmological time dilation and
$dV(z)/dz$ is the differential comoving volume, which can be
expressed as
\begin{equation}\label{comovingvolume}
\frac{dV(z)}{dz}=4\pi(\frac{c}{H_0})^3 (\int_0^z
\frac{dz}{\sqrt{1-\Omega_{\rm m}+\Omega_{\rm m}(1+z)^3}})^2
\frac{1}{\sqrt{1-\Omega_{\rm m}+\Omega_{\rm m}(1+z)^3}}.
\end{equation}

\section{Luminosity function and formation rate of GRBs}\label{sec:result}
In this section, we present results on the luminosity function and
cosmic formation rate of GRBs.

\subsection{Luminosity function}

\begin{figure}
  % Requires \usepackage{graphicx}
  \centering
  \includegraphics[width=0.8\textwidth]{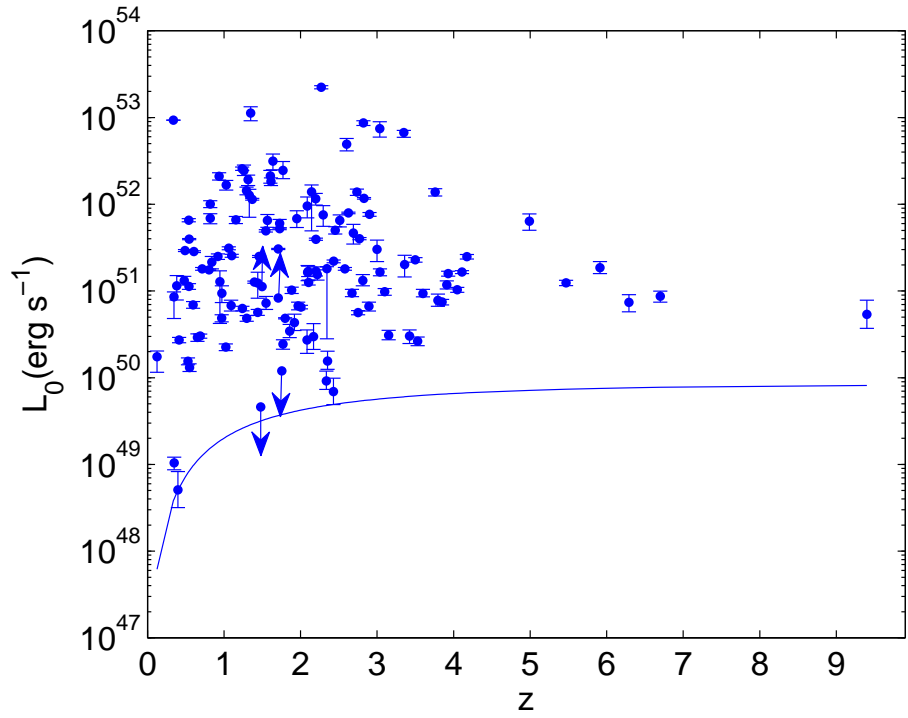}\\
  \caption{The non-evolving luminosity $L_0=L/(1+z)^{2.43}$ of 127 GRBs above the truncation line.
  The blue line represents observational limit. The error bars are $1\sigma$ errors.}\label{newdatafig}
\end{figure}

As discussed above, we get the form of luminosity evolution as
$g(z)=(1+z)^{2.43}$ by using the nonparametric $\tau$ test method.
The non-evolving luminosity $L_0$ is defined as $L_0 = L/g(z)$,
which is shown in Figure \ref{newdatafig}. Using this new data set,
the luminosity function $\psi(L_0)$ can be derived by using Lynden
Bell's $c^{-}$ method, which is shown in Figure
\ref{luminosityfunfig}. As shown in Figure \ref{luminosityfunfig},
the luminosity function $\psi(L_0)$ can be fitted with a broken
power law after removing the redshift evolution. The form of
luminosity function $\psi(L_0)$ for dimmer and brighter bursts are
fitted by
\begin{equation}\label{lumFunFit}
\psi(L_0)\propto\left\{
\begin{array}{ll}
        L_0^{-0.14\pm0.02} \ \
        & L_0 < L_0^{b}, \\
        L_0^{-0.70\pm0.03} \ \
        & L_0 > L_0^{b}, \\
\end{array}
\right.
\end{equation}
where $L_0^{b} = 1.43\times10^{51}~{\rm erg~s^{-1}}$ is the break
point. This result is consistent with previous work
\citep{Lloyd-Ronning2002,Yonetoku2004,Kocevski06,Wu2012}. It is
necessary to point out that this luminosity function is only the
present distribution at $z=0$ since the luminosity evolution is
removed. The luminosity function $\psi_z(L)$ at redshift $z$ will be
$\psi_z(L)=\psi(L/g(z))=\psi(L/(1+z)^{2.43})$. So the break
luminosity at $z$ is $L_z^b=L_0^b (1+z)^{2.43}$.

\subsection{Formation rate of GRBs}

Figure 5 presents the cumulative GRB formation rate $\phi(z)$.
According to equation (\ref{formationrate}), in order to get the
cosmic formation rate of GRBs, we need the differential form of
cumulative number distribution $d\phi(z)/dz$. Figure \ref{dphidzfig}
shows $(1+z) d\phi(z)/dz$ as function of redshift $z$. From this
figure, we find that $(1+z) d\phi(z)/dz$ increases quickly at $z <
1$, then keeps approximately constant for $1<z<4$, and decreases
sharply at $z>4$ with a power-law form. But we are more interested
in the comoving density rate. From equation (\ref{formationrate}),
we can calculate the GRBs formation rate $\rho(z)$, which is shown
in Figure \ref{formationratefig}. In Figure \ref{formationratefig},
the blue stepwise line represents the comoving cosmic formation of
GRBs as a function of redshift, and the error bar gives the
$1\sigma$ confidence level. The best-fitting power laws for
different segments are
\begin{equation}\label{formationratefit1}
\rho(z)\propto\left\{
\begin{array}{lll}
        (1+z)^{0.04\pm0.94} \ \
        & z<1, \\
        (1+z)^{-0.94\pm0.11} \ \
        & 1<z<4, \\
        (1+z)^{-4.36\pm0.48} \ \
        & z>4,
\end{array}
\right.
\end{equation}
with $95\%$ confidence level. From this equation, we can derive the
formation rate of GRBs at the local universe $\rho(0)=7.3\pm2.7~\rm
Gpc^{-3}~\rm yr^{-1}$, which is larger than previous studies, e.g.,
$\rho(0)\sim1.5 ~\rm Gpc^{-3}~\rm yr^{-1}$ \citep{Schmidt99},
$\rho(0)\sim0.5~\rm Gpc^{-3}~\rm yr^{-1}$ \citep{Guetta05}, and
$\rho(0)>0.5 ~\rm Gpc^{-3}~\rm yr^{-1}$ \citep{Pelangeon08}. The
main reason is that the GRB rate keeps constant at low redshift in
this paper, while it increases fast in other studies. But
\cite{Liang07} found that the rate of low-luminosity GRBs is
$\rho(0)\sim325 ~\rm Gpc^{-3}~\rm yr^{-1}$. This local rate is not
corrected for the jet beaming effect.

Obviously, the formation rate of GRBs is in contrast to previous
estimates of the comoving rate density by \citet{Yonetoku2004},
\citet{Kocevski06} and \citet{Wu2012} with the same method. Our
results shows that the formation rate of GRBs is almost constant at
$z<1.0$. But previous results give that the formation rate increases
quickly at $z < 1.0$ \citep{Yonetoku2004,Kocevski06,Wu2012}, which
is consistent with SFR observation \citep{Hopkins2006}. But our
result is consistent with that of \citet{Petrosian2009} well.
Interestingly, the evolution of $(1+z)d\phi(z)/dz$ shown in Figure
\ref{dphidzfig}, is consistent with the behavior of $\rho(z)$ in
\citet{Wu2012}. We also test our program with the same GRB data of
\citet{Yonetoku2004} and \citet{Wu2012}, and find that our
$(1+z)d\phi(z)/dz$ shows the same behaviors as the $\rho(z)$ in
\citet{Yonetoku2004} and \citet{Wu2012}. So they might omit the
$dV(z)/dz$ term in their calculations. Besides, \citet{Yonetoku2004}
and \citet{Wu2012} showed that the GRB rate increases as
$(1+z)^{6.0\pm1.4}$ and $(1+z)^{8.24\pm4.48}$ at $z<1$ respectively,
which are much quickly than $(1+z)^{2.4}$ of \citet{Kocevski06}.

Several comprehensive works studying the luminosity function and the
rate of long GRBs have been recently published using different
methods \citep[such as][]{Wanderman2010,Butler2010}. Assuming that
the luminosity function is redshift independent,
\citet{Wanderman2010} found that the power law index of luminosity
function is $0.22$ at low luminosity, and $1.4$ at high luminosity
with break $10^{52.5}$ erg~s$^{-1}$ using long GRBs with redshifts
determined from afterglow. The formation rate increases as
$(1+z)^{2.1}$ up to $z\sim3.0$ and it decreases as $(1+z)^{-1.4}$ at
$z>3.0$. \citet{Butler2010} derived that the luminosity function is
nearly flat $\propto L^{-0.2}$ below break $10^{52.7}$ erg~s$^{-1}$,
and declines $\propto L^{-3.0}$ using a large sample of GRBs
detected by \emph{Swift}. The GRB rate is similar as that of
\citet{Wanderman2010}. These results are different from our results.
One reason is the GRB sample. We use the latest GRB sample, which
have redshift observed by \emph{Swift} and spectral parameters given
by \emph{Fermi-GBM} and \emph{Konus-wind}. The luminosity function
evolution may be the most important reason. \citet{Wanderman2010}
assumed no redshift evolution of luminosity function. In the fitting
of \citet{Butler2010}, no luminosity evolution is required to
produce the observed number of GRBs. But strong evolution of
luminosity function is found in literatures. The evolution of
luminosity can be parameterized as $(1+z)^{1.4}$ (Lloyd-Ronning et
al. 2002). \citet{Yonetoku2004} and \citet{Wu2012} found that the
evolution factor is $g(z)=(1+z)^{2.60}$ and $g(z)=(1+z)^{2.30}$
respectively. \citet{Tan14} found that the luminosity function of
GRB evolves with a redshift-dependent break luminosity
$L_b=1.2\times10^{51}(1+z)^2 \rm~erg~s^{-1}$, which is similar with
our result. \citet{Virgili11} found that a evolution factor
$(1+z)^{1.0\pm0.2}$ of luminosity function can fit the BATSE and
Swift data. These works suggest that take a evolution factor into
consideration is necessary. Besides, our GRB sample including those
GRBs dimmer than $10^{51}\rm~erg~s^{-1}$ is another important
reason. For example, \citet{Kistler2008} found that the density of
GRB is much higher at $z<1$ if they included GRBs dimmer than
$10^{51}\rm~erg~s^{-1}$. In this work, we use Lynden-Bell $c^-$
method to correct the data truncated effect and consider the
evolution of GRB luminosity function. So we don't need to omit these
dim GRBs.
%\citep[i.e.,][]{Lloyd-Ronning2002,Wei03,Yonetoku2004,Kocevski06,Cao11,Virgili11,Tan14}.

The relation between SFR and formation rate of GRBs is attractive.
We also compare our result with the observed SFR from
\citet{Hopkins2006} in Figure \ref{SFRfig}. Obviously, it is
consistent with the observed SFR at $z>1.0$, but remarkably
different at $z<1.0$. This trend means that the formation rate of
GRBs $\rho(z)$ does not trace SFR at low redshift $z<1.0$. But at
high redshift $z>4.0$, our result is consistent with the SFR derived
from GRBs \citep{Yuksel2008,Wang09,Kistler2009,Wang13}. This result
is different with others in previous literatures
\citep{Lloyd-Ronning2002,Yonetoku2004,Kocevski06,Wu2012}, but it is
consistent with that of \citet{Petrosian2009}. There are also some
previous works shown that the long GRBs may not unbiased tracers of
SFR at low redshift. A strong dependence of the GRB rate on
host-galaxy properties out to $z\sim 1.0$ is found by
\cite{Perley2013}. So use GRBs as direct tracers of the cosmic SFR
is cautious at $z<1.0$ \citep{Perley2013}. \cite{Vergani2014} found
that the mass distribution of long GRB host galaxy is different with
the expected from star-forming galaxies observed in deep survey,
which suggests that long GRBs are not unbiased tracers of star
formation activity at least at $z<1.0$. They also found that long
GRB rate can directly trace the SFR starting from $z\sim4$ and
above.

\section{Testing with Monte Carlo simulation}
In this section, we use Monte Carlo simulation to test our results.
Firstly, we simulate a set of data $(L_0,z)$ which follows the
distribution described by equation (\ref{lumFunFit}) and equation
(\ref{formationratefit1}) using Monte Carlo method. Then, we
transfer the luminosity $L_0$ to $L$ through $L=L_0(1+z)^k$, where
$k=2.43$. So we can get sets of pseudo data of GRB luminosity and
redshift $(L,z)$. In the simulations, we create 200 pseudo samples.
Each sample contains 130 GRBs. Then we use Lynden-Bell $c^-$ method
and nonparametric $\tau$ test method to calculate the distributions
of these pseudo samples. Finally, we compare the simulated data with
observed data.

Figure \ref{CompareDatafig} shows the comparing results. The four
panels give the luminosity-redshift distribution, luminosity
function, cumulative distribution and $\log N-\log S$ distribution.
In the panel a, we randomly choose one pseudo sample of GRB from the
200 samples to compare with the observed data. The red dots and the
blue dots represent the observed data and the simulated data,
respectively. From this panel, we can see that the simulated data
and the observed data have similar distributions. The other three
panels b, c and d show the comparisons of the luminosity function,
cumulative distribution and $\log N-\log S$ distribution between the
observed data and mock data. The red curves show the distributions
of the observed data, blue curves give the distributions of all of
the 200 pseudo samples of GRB data and the blue curves are the mean
distributions of the 200 pseudo samples. We perform the
Kolmogorov-Smirnov test between observed data and the mean
distributions of simulated data. The chance probabilities of the
three tests are 0.49, 0.86 and 0.96, respectively. From these
panels, we can also conclude that the distribution of the observed
data lie in the region of those pseudo data, which means that the
derived luminosity function and formation rate of GRBs are correct.

In order to test whether long GRBs are unbiased tracers of SFR at
low redshift, we simulate 200 new pseudo samples of GRBs by assuming
that the GRB rate follows the SFR from \cite{Yuksel2008}, i.e.,
$\rho(z)\propto(1+z)^{3.4}$ at $z<1$, $\rho(z)\propto(1+z)^{-0.3}$
at $1<z<4$ and $\rho(z)\propto(1+z)^{-3.5}$ at $z>4$. Then we use
the same method to calculate the distributions of these pseudo data.
We find that the cumulative redshift distribution of observed data
is not consistent with the pseudo data, which is shown in Figure
\ref{CompareSFR}. The red, blue and green curves have the same
meanings as those in Figure \ref{CompareDatafig}. From Figure
\ref{CompareSFR}, we can see that part of the cumulative redshift
distribution line of observed data lies outside of the region
occupied by pseudo GRB data, especially at $z<1.0$. The
Kolmogorov-Smirnov test between the distribution from observed data
and the mean distribution of simulated data gives the chance
probability of $p=6.9\times10^{-12}$. It means that long GRBs are
not direct tracers of SFR at $z<1.0$.

\section{Conclusions and Discussions}\label{sec:conclusion}
\begin{figure}
\centering
  \includegraphics[width=0.8\textwidth]{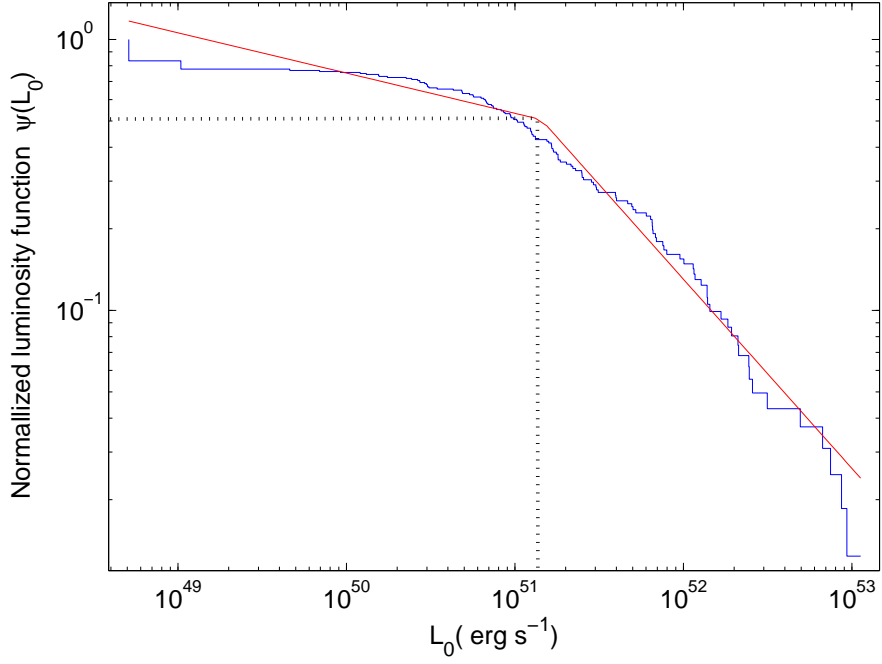}\\
  \caption{The cumulative luminosity function $\psi(L_0)$, which
  is normalized to unity at the lowest luminosity. The red line is
  the best fit with a broken power law model. The luminosity
function can be expressed as $\psi(L_0)\propto L_0^{-0.14\pm0.02}$
for dim GRBs and $\psi(L_0)\propto L_0^{-0.70\pm0.03}$ for bright
GRBs, with the break point $L_{0}^{b}=1.43\times10^{51}~{\rm
erg~s^{-1}}$.}\label{luminosityfunfig}
\end{figure}

\begin{figure}
\centering
  % Requires \usepackage{graphicx}
  \includegraphics[width=0.8\textwidth]{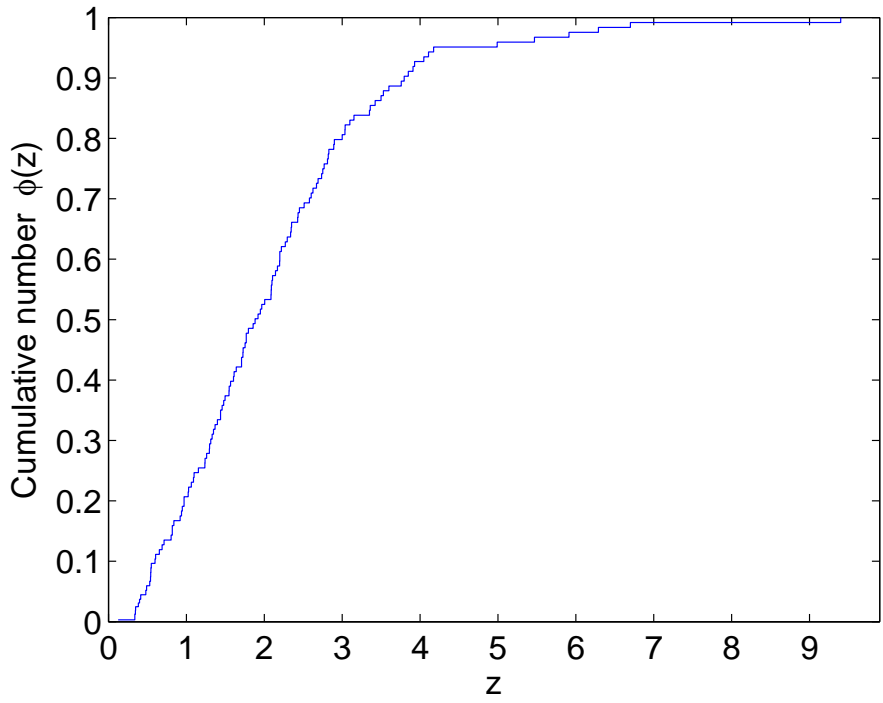}\\
  \caption{The normalized cumulative redshift distribution of GRBs.}\label{densityfunfig}
\end{figure}

In this paper, we use Lynden-Bell's $c^-$ method to study the
luminosity function and formation rate of \emph{Swift} long GRBs
without any assumptions. First, we use a $\tau$ statistical method
to separate the luminosity evolution from the stable form of the
luminosity function by choosing the evolution form $g(z)=(1+z)^k$.
The most proper $k$ is $k=2.43_{-0.38}^{+0.41}$, which gives
$\tau=0$. This value is similar with those of \cite{Yonetoku2004},
\cite{Wu2012} and \citet{Kocevski06}. After correcting the
luminosity evolution by $L_0 = L/(1+z)^{2.43}$, the cumulative
luminosity function $\psi(L_0)$ and cumulative number distribution
$\phi(z)$ of GRBs can be calculated, which are shown in Figure
\ref{luminosityfunfig} and Figure \ref{densityfunfig}. The
luminosity function of GRBs can be well fitted with a broken power
law form as $\psi(L_0)\propto L_0^{-0.14\pm0.02}$ and
$\psi(L_0)\propto L_0^{-0.70\pm0.03}$ for $L_0 < L_0^{\rm b}$ and
$L_0 > L_0^{\rm b}$ respectively, where $L_0^{\rm
b}=1.43\times10^{51}~{\rm erg~s^{-1}}$ is the break point.

\begin{figure}
\centering
  \includegraphics[width=0.8\textwidth]{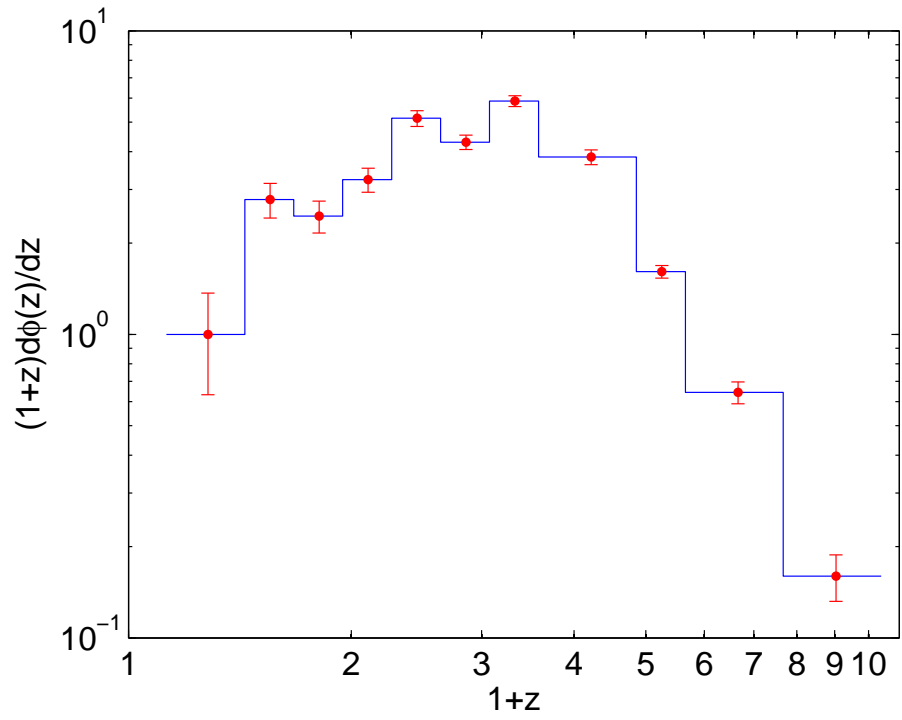}\\
  \caption{The evolution of $(1+z)\frac{d\phi(z)}{dz}$ as function of redshift $z$ with $1\sigma$ errors, which is
  normalized to unity at the first point.}\label{dphidzfig}
\end{figure}

\begin{figure}
\centering
  % Requires \usepackage{graphicx}
  \centering
  \includegraphics[width=0.8\textwidth]{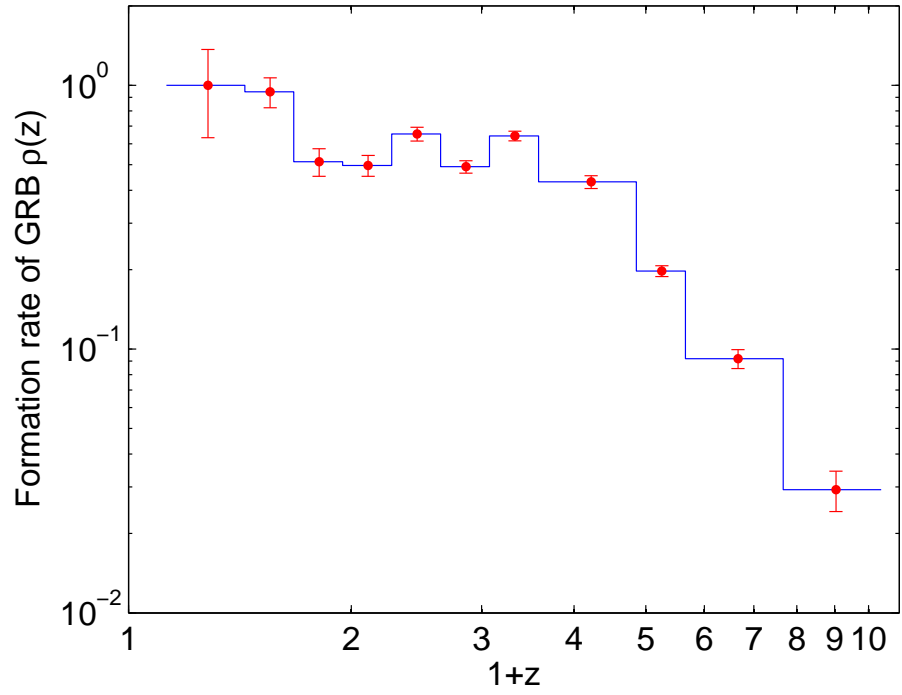}\\
  \caption{The comoving formation rate $\rho(z)$ of GRBs, which is normalized to unity
  at the first point. The $1\sigma$ error is also shown.}\label{formationratefig}
\end{figure}

\begin{figure}
\centering
  \includegraphics[width=0.8\textwidth]{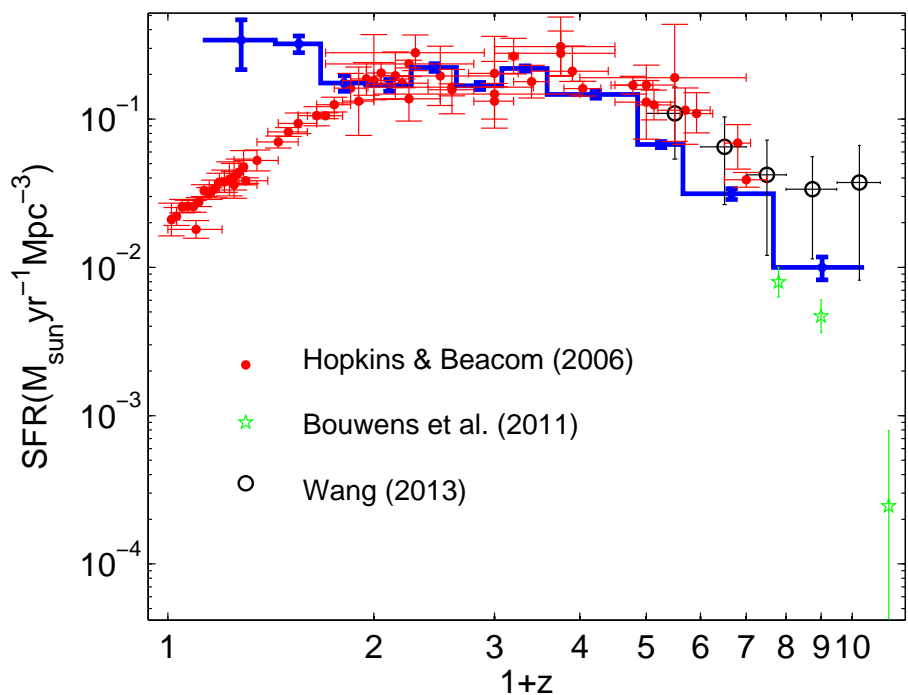}\\
  \caption{The comparison between GRB formation rate $\rho(z)$ (blue) and the observed SFR. The SFR data are taken from Hopkins \& Beacom
  (2006),  which are shown as red dots. The SFR data from \cite{Bouwens2011}
  (stars) and \cite{Wang13} (open circles) are also used. All error bars are $1\sigma$ errors.}\label{SFRfig}
\end{figure}

\begin{figure}
%\centering
 \includegraphics[width=0.5\textwidth]{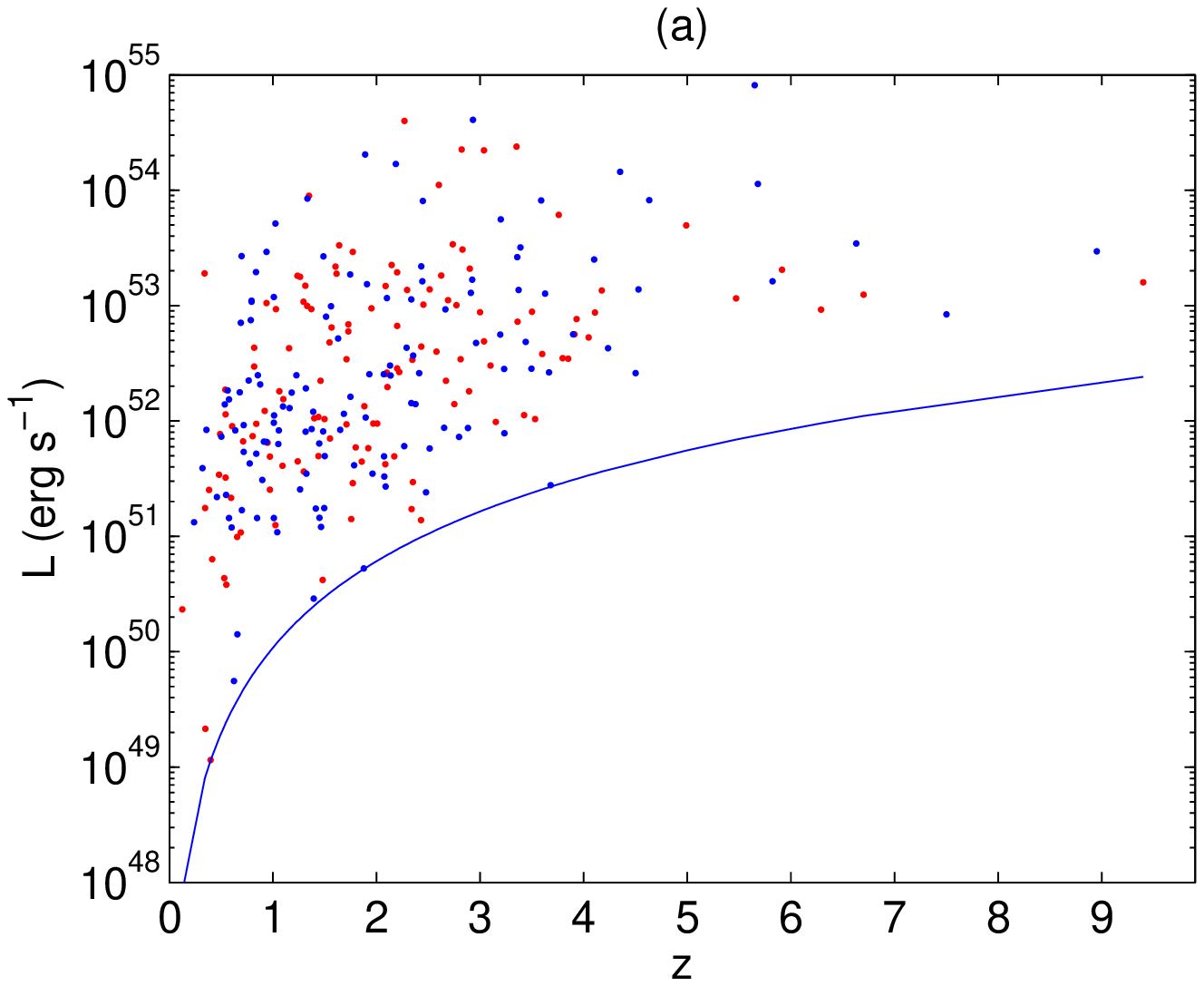}
 \includegraphics[width=0.5\textwidth]{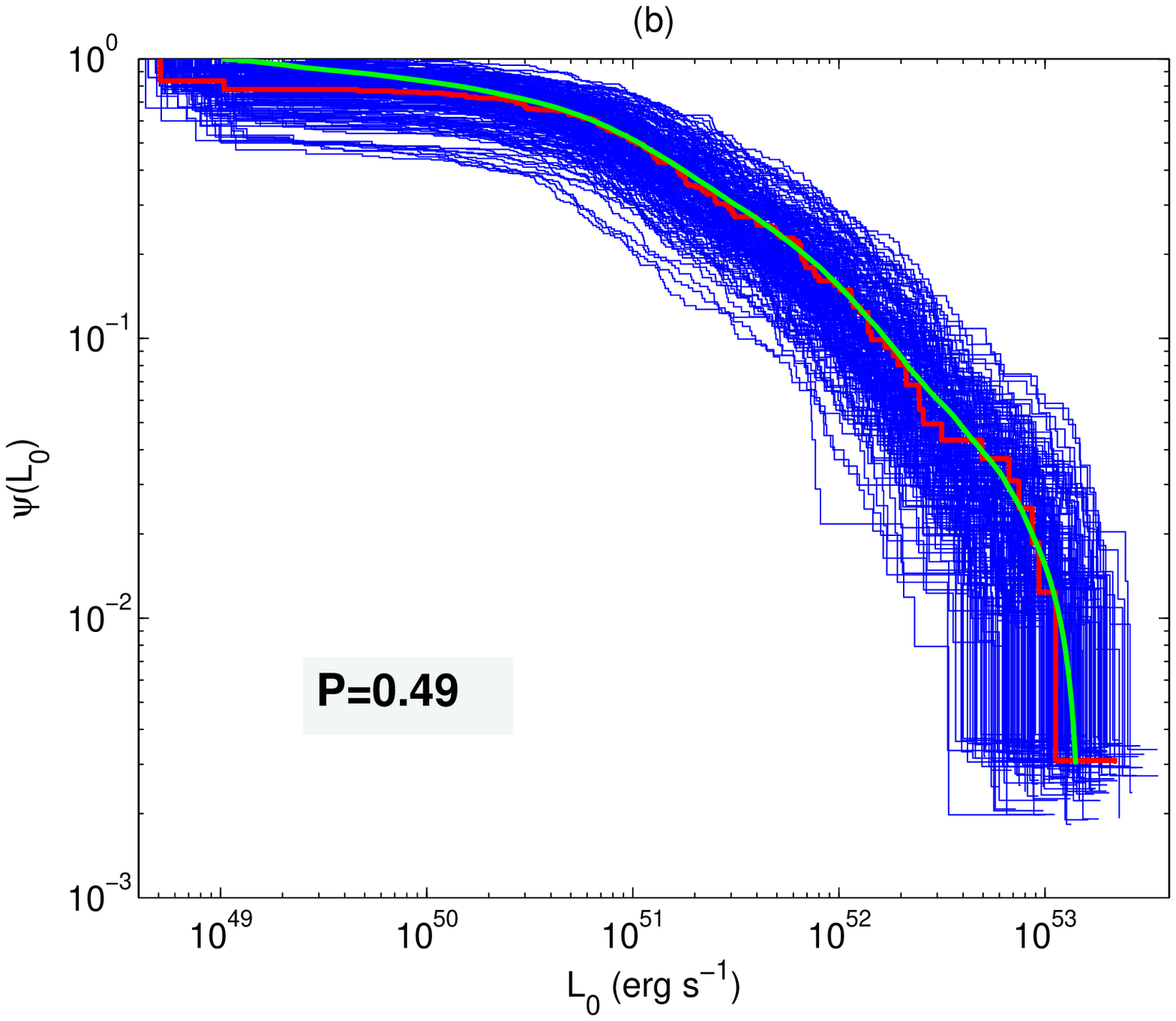}
 \includegraphics[width=0.5\textwidth]{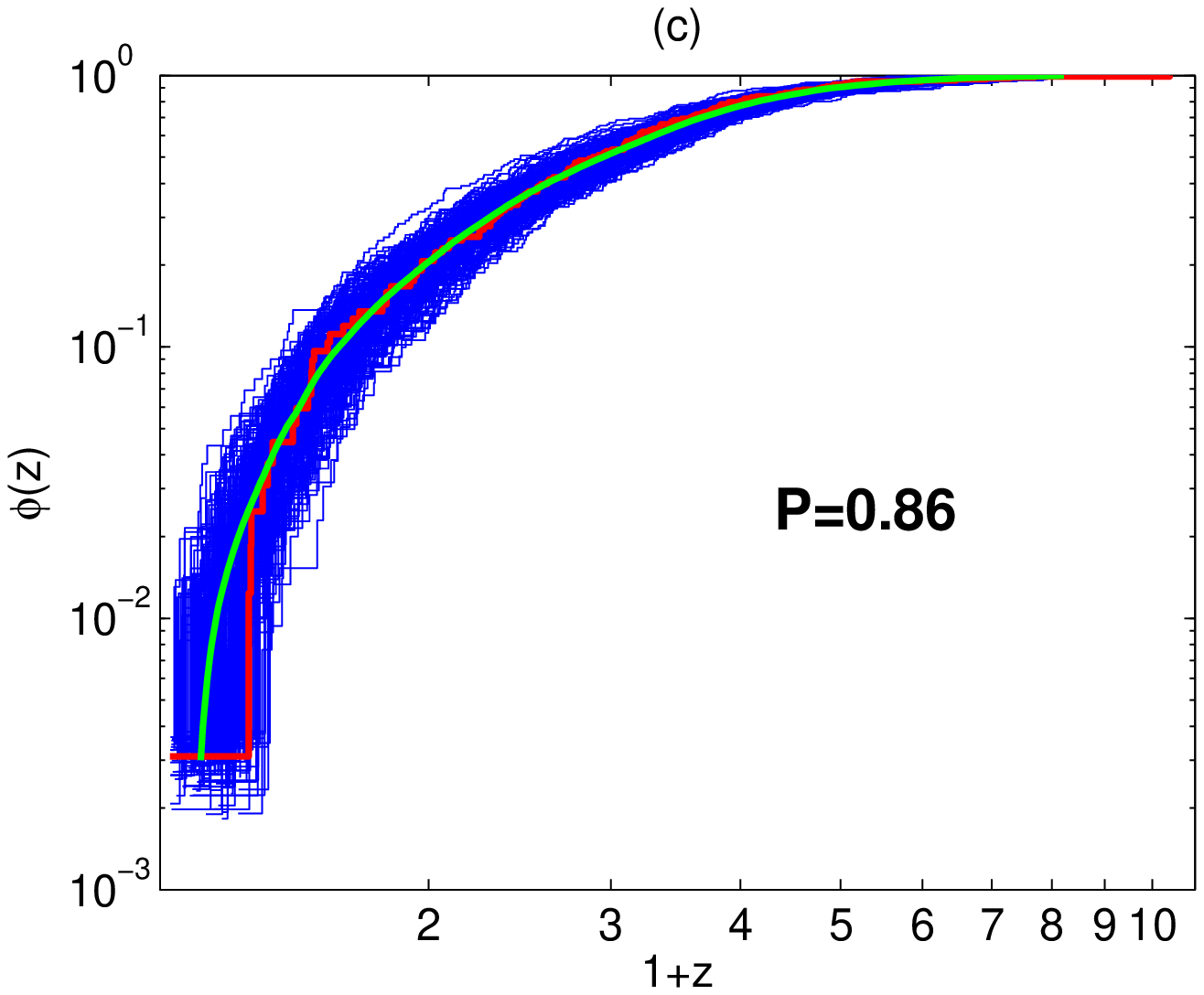}
 \includegraphics[width=0.5\textwidth]{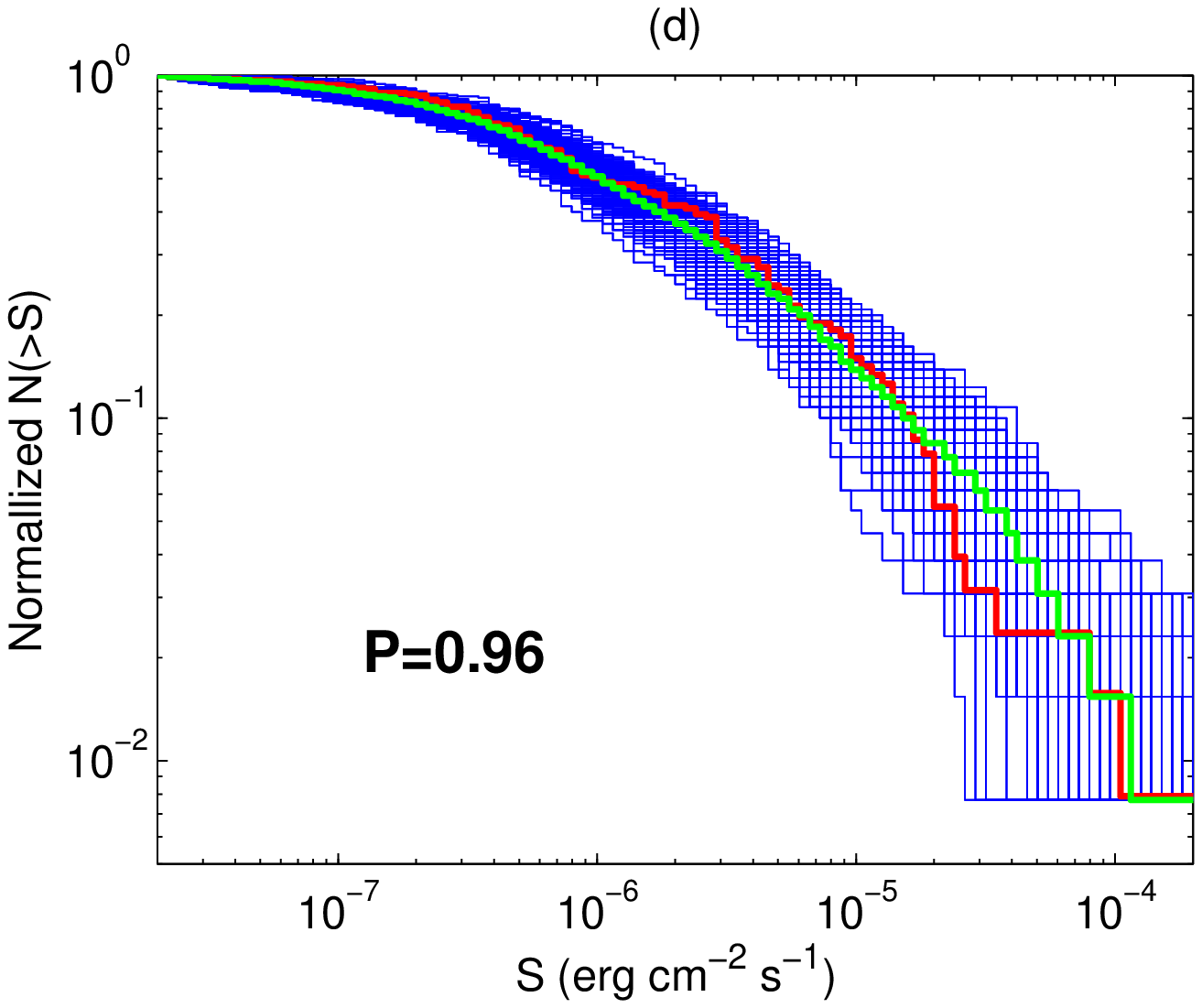}
 \caption{The comparison of the simulated data (blue)
 and the observed data (red). These four panels show luminosity-redshift distribution, cumulative luminosity function, cumulative number
 distribution and $\log N-\log S$ distribution respectively. For panel (a), we choose one sample from the 200 simulated samples randomly.
 For other panels (b), (c) and (d), the green curves represent the mean distribution of those 200 simulated samples. The chance probabilities of
 Kolmogorov-Smirnov tests between the distributions of observed data and the mean distributions of the simulated data are also presented.}\label{CompareDatafig}
\end{figure}

\begin{figure}
\centering
  \includegraphics[width=0.8\textwidth]{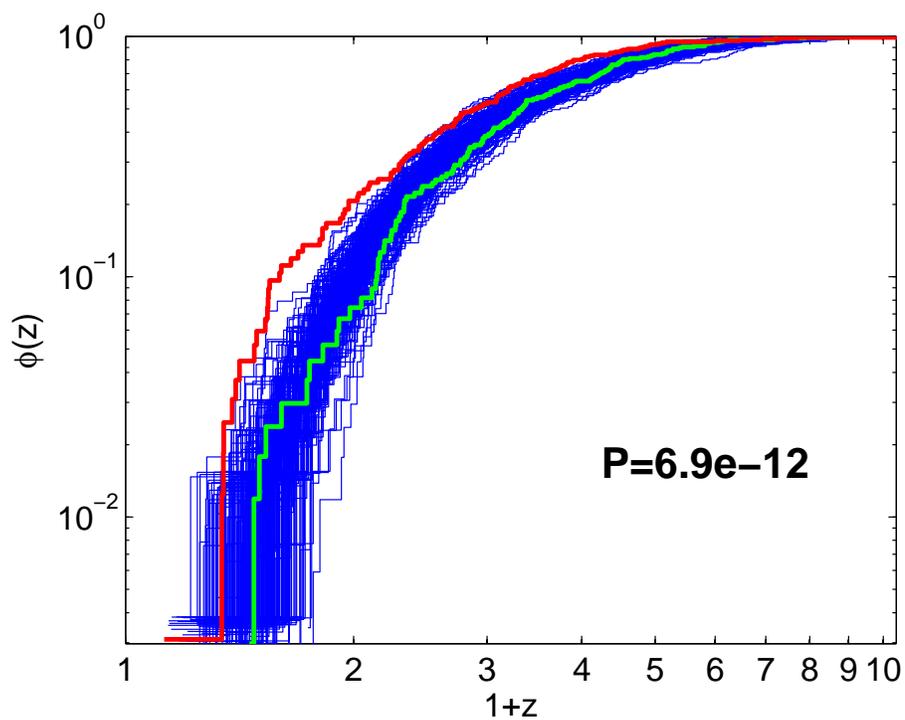}\\
  \caption{The comparison between the cumulative redshift distributions of simulated data (blue) and observed data (red).
  The mean distribution of the 200 simulated samples is given by green curve.
  In this case we use the SFR from \cite{Yuksel2008}. The chance probability of the
  Kolmogorov-Smirnov test between the observed data and the
  simulated data is $p=6.9\times10^{-12}$, from which the observed data and the simulated data from the same sample can be discarded.}\label{CompareSFR}
\end{figure}

We also derive the formation rate of GRBs through the differential
form of the cumulative number distribution $\phi(z)$. Figure
\ref{dphidzfig} shows the evolution of $(1+z)\frac{d\phi(z)}{dz}$.
We find that $(1+z)\frac{d\phi(z)}{dz}$ increases quickly at $z<1$,
then remains roughly constant at $1<z<4$ and finally decreases
rapidly at high redshift. From equation (\ref{formationrate}), the
cosmic formation rate of GRBs $\rho(z)$ is derived, which is shown
in Figure \ref{formationratefig}. The best-fitting power laws for
different redshift segments are $\rho(z)\propto(1+z)^{0.04}$,
$\rho(z)\propto(1+z)^{-0.94}$, $\rho(z)\propto(1+z)^{-4.36}$ for
$z<1.0$, $1.0<z<4.0$ and $z>4.0$ respectively. Our results show that
the formation rate of GRBs is almost constant at $z<1.0$. But
previous results give that the formation rate increases quickly at
$z < 1.0$ \citep{Yonetoku2004,Kocevski06,Wu2012}. But
\citet{Yonetoku2004} and \citet{Kocevski06} used the pseudo
redsihfts of GRBs rather than the observed redshifts. Besides, we
find the $\rho(z)$ in \cite{Wu2012} and \cite{Yonetoku2004} increase
fast at $z<1.0$, which has the similar behavior of
$(1+z)\frac{d\phi(z)}{dz}$ shown in Figure \ref{dphidzfig}. So they
might omit the $dV(z)/dz$ term in their calculations. From Figure
\ref{SFRfig}, it is easily to find that GRB formation rate $\rho(z)$
is consistent with observed SFR at $z>1.0$ but entirely different at
$z<1.0$. It means that the formation rate of GRBs only traces SFR at
$z>1.0$, which is different with previous work
\citep{Yonetoku2004,Kocevski06,Wu2012}. We find the low-redshift
excess of GRB rate for the first time.

Surprisingly, we find that formation rate of GRBs is consistent with
SFR at $z>1.0$, but shows an excess at low redshift $z<1.0$ for the
first time, which is different with previous works. Our result shows
that formation rate of GRBs is larger than SFR at $z<1.0$. Below, we
will discuss some possible reasons for this low-redshift excess.

The first one is that the definition of long GRBs is not clear. In
classical method, the long GRBs are defined by $T_{90}>2$ s
\citep{Kou93}. There is no clear boundary line in this diagram to
separate the long and short GRBs. Moreover, $T_{90}$ is an observed
time scale, which represents different time for GRBs at different
redshifts. Meanwhile, the observations of low-redshift long GRBs,
such as GRB 060614 at $z=0.125$ and GRB 060505 at $z=0.089$, show no
association of supernovae \citep{Gal-Yam06,Fynbo06,Gehrels06}. So
more physical criterions are required to classify GRBs. Because only
a subclass of GRBs can trace the SFR. Some attempts have been
performed \citep{Zhang06,ZhangB07,Zhang09,Bloom08,Bromberg13,Lv14}.
It has been suggested that GRBs can be classified physically into
Type I (compact star origin) and Type II (massive star origin)
\citep{Zhang06,ZhangB07}.

The second one is that some selection effects have not been included
in analysis. For example, it is easier to measure the redshift of
those GRBs which are in lower redshift and therefore create a bias
toward low redshift GRBs. It means that we lose some high redshift
GRBs, so the formation rate of GRBs at low redshift we calculated
will larger than the SFR. This bias can be removed by using samples
with high completeness in the GRB redshift measurements. There are
three of such a sample in previous literature
\citep{Greiner2011,Salvaterra2012,Hjorth2012}. It could be
considered in the future works.

The third one is that there may exist a subclass GRBs, i.e.,
low-luminosity GRBs \citep{Cobb06,Pian06,Soderberg06,Liang07}. The
local rate of low-luminosity GRBs may be high, i.e.,
$\rho(0)=100-1000~\rm yr^{-1}~\rm
Gpc^{-3}$\citep{Soderberg06,Liang07}, much higher than
high-luminosity GRBs. The progenitors of low-luminosity GRBs may be
different with those of high-luminosity GRBs
\citep{Mazzali06,Soderberg06}. The contamination from low-luminosity
GRBs could lead to the low-redshift excess.

\acknowledgments We thank the referee for helpful commentary on the
manuscript. We acknowledge the use of public data from the
\emph{Swift} data archive. We thank Bing Zhang, Yun-Wei Yu, Shi-Wei
Wu and Jin-Jun Geng for helpful discussions. This work is supported
by the National Basic Research Program of China (973 Program, grant
No. 2014CB845800), the National Natural Science Foundation of China
(grants 11422325, 11373022, 11033002, and J1210039), the Excellent
Youth Foundation of Jiangsu Province (BK20140016), and the Program
for New Century Excellent Talents in University (grant No.
NCET-13-0279). KSC is supported by the CRF Grants of the Government
of the Hong Kong SAR under HUKST4/CRF/13G.

\begin{deluxetable}{lllllllll}\label{GRBdata}
\tablecolumns{9} \tablewidth{0pc} \tabletypesize{\scriptsize}
\tablecaption{List of long GRBs used in this paper. It gives the name, redshift $z$, spectra parameters $\alpha$ \& $\beta$, rest frame peak energy $E_{p}$, peak flux $F$, energy range, bolometric luminosity $L$ in $1-10^4~\rm keV$ and reference of the parameters of spectrum of each GRB.}
\tablehead{ \colhead{GRB} &
\colhead{z} & \colhead{$\alpha$} & \colhead{$\beta^b$} &
\colhead{$\rm E_p(keV)$} & \colhead{Flux$\rm(erg/cm^{2}/s)$} & \colhead{Range\rm(keV)} &
\colhead{$L(\rm erg/s)$} & \colhead{Ref.}} \startdata

050318  &   1.44    & $  -1.34  ^{+ 0.32    }_{-    0.32    }$ &    ...                 & $ 63.52   ^{+ 11.07   }_{-    11.07   }$  &$( 2.2 \pm 0.17    )\times10^{ -7  }$              &   15  -   150 & $ 4.96    ^{+ 0.38    }_{-    0.38    }\times10^{ 51  }$  &   1   \\
050401  &   2.9 & $ -0.83   ^{+ 0.21    }_{-    0.21    }$ & $  -2.37   ^{+ 0.14    }_{-    0.14    }$ & $  119 ^{+ 26  }_{-    26  }$  &$( 2.45    \pm 0.12    )\times10^{ -6  }$              &   20  -   2000    & $ 2.09    ^{+ 0.10    }_{-    0.10    }\times10^{ 53  }$  &   2   \\
050416A &   0.6535  & $  -1.0                   $ & -3.4                    & $ 15.73   ^{+ 2.42    }_{-    2.42    }$  &$  5   \pm 0.5                     ^a$ &   15  -   150 & $ 9.89    ^{+ 0.99    }_{-    0.99    }\times10^{ 50  }$  &   1   \\
050525  &   0.606   & $  -0.99  ^{+ 0.11    }_{-    0.11    }$ &    ...                 & $ 79.08   ^{+ 3.74    }_{-    3.74    }$  &$  47.7    \pm 1.2                     ^a$ &   15  -   150 & $ 9.00    ^{+ 0.23    }_{-    0.23    }\times10^{ 51  }$  &   1   \\
050603  &   2.821   & $ -0.79   ^{+ 0.06    }_{-    0.06    }$ & $  -2.15   ^{+ 0.09    }_{-    0.09    }$ & $  349 ^{+ 28  }_{-    28  }$  &$( 3.2 \pm 0.2 )\times10^{ -5  }$              &   20  -   3000    & $ 2.25    ^{+ 0.14    }_{-    0.14    }\times10^{ 54  }$  &   3   \\
050802  &   1.71    & $  -1.6   ^{+ 0.1 }_{-    0.1 }$ &    ...                 & $ >70.59                  $   &$( 2.21    \pm 3.53    )\times10^{ -7  }$              &   15  -   150 & $ >9.34                   \times10^{  51  }$  &   1   \\
050904  &   6.29    & $ -1.15   ^{+ 0.12    }_{-    0.12    }$ &    ...                 & $ 314 ^{+ 173 }_{-    89  }$  &$( 1.84    \pm 0.41    )\times10^{ -7  }$              &   15  -   5000    & $ 9.25    ^{+ 2.06    }_{-    2.06    }\times10^{ 52  }$  &   4   \\
050922C &   2.198   & $  -0.83  ^{+ 0.24    }_{-    0.24    }$ &    ...                 & $ 130.8   ^{+ 36.9    }_{-    36.9    }$  &$  4.5 ^{+ 0.72    }_{-    1.53    }\times10^{ -6  }$      &   20  -   2000    & $ 1.95    ^{+ 0.30    }_{-    0.30    }\times10^{ 53  }$  &   1   \\
051001  &   2.4296  & $ -1.12   ^{+ 0.66    }_{-    0.56    }$ &    ...                 & $ 44.38   ^{+ 11.48   }_{-    11.48   }$  &$  1.95    ^{+ 0.83    }_{-    0.57    }\times10^{ -8  }$      &   15  -   350 & $ 1.38    ^{+ 0.59    }_{-    0.40    }\times10^{ 51  }$  &   5   \\
051109A &   2.346   & $ -1.25   ^{+ 0.44    }_{-    0.59    }$ &    ...                 & $ 161 ^{+ 224 }_{-    58  }$  &$  5.8 ^{+ 0.3 }_{-    4.9 }\times10^{ -7  }$      &   20  -   500 & $ 3.40    ^{+ 0.18    }_{-    2.87    }\times10^{ 52  }$  &   6   \\
051111  &   1.55    & $ -0.98   ^{+ 0.25    }_{-    0.24    }$ &    ...                 & $ 179.70  ^{+ 316.76  }_{-    54.52   }$  &$  3.41    ^{+ 0.66    }_{-    0.52    }\times10^{ -7  }$      &   15  -   350 & $ 7.04    ^{+ 1.36    }_{-    1.07    }\times10^{ 51  }$  &   5   \\
060115  &   3.53    & $ -1.0    ^{+ 0.5 }_{-    0.5 }$ &    ...                 & $ 62  ^{+ 31  }_{-    10  }$  &$  0.9 \pm 0.1                     ^a$ &   15  -   150 & $ 1.04    ^{+ 0.12    }_{-    0.12    }\times10^{ 52  }$  &   7   \\
060124  &   2.296   & $ -1.29   ^{+ 0.14    }_{-    0.11    }$ & $  -2.25   ^{+ 0.27    }_{-    0.88    }$ & $  247.76  ^{+ 130.91  }_{-    88.75   }$  &$  2.66    ^{+ 0.74    }_{-    0.69    }\times10^{ -6  }$      &   20  -   2000    & $ 1.37    ^{+ 0.38    }_{-    0.35    }\times10^{ 53  }$  &   8   \\
060206  &   4.048   & $  -1.12  ^{+ 0.3 }_{-    0.3 }$ &    ...                 & $ 81  ^{+ 22  }_{-    22  }$  &$( 2.02    \pm 0.13    )\times10^{ -7  }$              &   15  -   150 & $ 5.29    ^{+ 0.34    }_{-    0.34    }\times10^{ 52  }$  &   1   \\
060210  &   3.91    & $  -1.12  ^{+ 0.26    }_{-    0.26    }$ &    ...                 & $ 117 ^{+ 23  }_{-    23  }$  &$  2.8 \pm 0.3                     ^a$ &   15  -   150 & $ 5.64    ^{+ 0.60    }_{-    0.60    }\times10^{ 52  }$  &   1   \\
060306  &   3.5 & $  -1.2   ^{+ 0.5 }_{-    0.5 }$ &    ...                 & $ 70  ^{+ 18  }_{-    18  }$  &$( 4.71    \pm 0.278   )\times10^{ -7  }$              &   15  -   150 & $ 8.85    ^{+ 0.51    }_{-    0.51    }\times10^{ 52  }$  &   1   \\
060428B &   0.348   & $ -0.94   ^{+ 1.30    }_{-    1.30    }$ &    ...                 & $ 21.7    ^{+ 14  }_{-    14  }$  &$  0.6 \pm 0.1                     ^a$ &   15  -   150 & $ 2.15    ^{+ 0.36    }_{-    0.36    }\times10^{ 49  }$  &   9   \\
060614  &   0.125   & $ -1.57   ^{+ 0.12    }_{-    0.14    }$ &    ...                 & $ 302 ^{+ 214 }_{-    85  }$  &$( 4.5 \pm 0.7 )\times10^{ -6  }$              &   20  -   2000    & $ 2.33    ^{+ 0.37    }_{-    0.79    }\times10^{ 50  }$  &   10  \\
060707  &   3.425   & $ -0.66   ^{+ 0.63    }_{-    0.63    }$ &    ...                 & $ 66  ^{+ 25  }_{-    10  }$  &$  1.1 \pm 0.2                     ^a$ &   15  -   150 & $ 1.12    ^{+ 0.20    }_{-    0.20    }\times10^{ 52  }$  &   11  \\
060708  &   1.92    & $ -0.93   ^{+ 0.47    }_{-    0.43    }$ &    ...                 & $ 87.45   ^{+ 83.35   }_{-    18.94   }$  &$  1.78    ^{+ 0.45    }_{-    0.33    }\times10^{ -7  }$      &   15  -   350 & $ 5.82    ^{+ 1.47    }_{-    1.08    }\times10^{ 51  }$  &   5   \\
060814  &   0.84    & $ -1.43   ^{+ 0.16    }_{-    0.16    }$ &    ...                 & $ 257 ^{+ 122 }_{-    58  }$  &$( 2.13    \pm 0.35    )\times10^{ -6  }$              &   20  -   1000    & $ 9.46    ^{+ 1.55    }_{-    1.55    }\times10^{ 51  }$  &   12  \\
060908  &   1.8836  & $  -0.93  ^{+ 0.25    }_{-    0.25    }$ &    ...                 & $ 148 ^{+ 72  }_{-    72  }$  &$( 2.81    \pm 0.23    )\times10^{ -7  }$              &   15  -   150 & $ 1.34    ^{+ 0.11    }_{-    0.11    }\times10^{ 52  }$  &   1   \\
060927  &   5.47    & $  -0.81  ^{+ 0.36    }_{-    0.36    }$ &    ...                 & $ 71  ^{+ 14  }_{-    14  }$  &$( 2.47    \pm 0.17    )\times10^{ -7  }$              &   15  -   150 & $ 1.16    ^{+ 0.08    }_{-    0.08    }\times10^{ 53  }$  &   1   \\
061007  &   1.261   & $ -0.53   ^{+ 0.09    }_{-    0.08    }$ & $  -2.61   ^{+ 0.25    }_{-    0.49    }$ & $  498 ^{+ 54  }_{-    48  }$  &$  1.95    ^{+ 0.31    }_{-    0.24    }\times10^{ -5  }$      &   20  -   10000   & $ 1.78    ^{+ 0.28    }_{-    0.22    }\times10^{ 53  }$  &   13  \\
061021  &   0.3463  & $  -1.22  ^{+ 0.12    }_{-    0.14    }$ &    ...                 & $ 777 ^{+ 549 }_{-    237 }$  &$  3.72    ^{+ 0.53    }_{-    1.62    }\times10^{ -6  }$      &   20  -   2000    & $ 1.76    ^{+ 0.25    }_{-    0.77    }\times10^{ 51  }$  &   14  \\
061121  &   1.314   & $  -1.32  ^{+ 0.04    }_{-    0.05    }$ &    ...                 & $ 606 ^{+ 90  }_{-    72  }$  &$  1.28    ^{+ 0.16    }_{-    0.19    }\times10^{ -5  }$      &   20  -   5000    & $ 1.48    ^{+ 0.19    }_{-    0.22    }\times10^{ 53  }$  &   15  \\
061222A &   2.088   & $ -1.00   ^{+ 0.05    }_{-    0.05    }$ & $  -2.32   ^{+ 0.38    }_{-    0.38    }$ & $  353 ^{+ 54  }_{-    54  }$  &$( 4.8 \pm 1.3 )\times10^{ -6  }$              &   20  -   10000   & $ 1.48    ^{+ 0.40    }_{-    0.40    }\times10^{ 53  }$  &   1   \\
070110  &   2.352   & $ -1.15   ^{+ 0.45    }_{-    0.41    }$ &    ...                 & $ 108.33  ^{+ 183.02  }_{-    16.29   }$  &$( 5.168   \pm 0.831   )\times10^{ -6  }$              &   15  -   350 & $ 2.95    ^{+ 0.87    }_{-    0.59    }\times10^{ 51  }$  &   5   \\
070129  &   2.3384  & $ -1.33   ^{+ 0.68    }_{-    0.59    }$ &    ...                 & $ 65.96   ^{+ 179.79  }_{-    63.48   }$  &$  2.72    ^{+ 0.8 }_{-    0.55    }\times10^{ -8  }$      &   15  -   350 & $ 1.72    ^{+ 0.51    }_{-    0.35    }\times10^{ 51  }$  &   5   \\
070306  &   1.497   & $  -1.67  ^{+ 0.1 }_{-    0.1 }$ &    ...                 & $ >105                    $   &$( 3.04    \pm 0.164   )\times10^{ -7  }$              &   15  -   150 & $ >1.04                   \times10^{  52  }$  &   1   \\
070508  &   0.82    & $ -0.81   ^{+ 0.07    }_{-    0.07    }$ &    ...                 & $ 188 ^{+ 8   }_{-    8   }$  &$  8.3 ^{+ 1.03    }_{-    1.11    }\times10^{ -6  }$      &   20  -   1000    & $ 2.96    ^{+ 0.37    }_{-    0.40    }\times10^{ 52  }$  &   16  \\
070714B &   0.92    & $ -0.86   ^{+ 0.1 }_{-    0.1 }$ &    ...                 & $ 1120    ^{+ 780 }_{-    380 }$  &$  2.7 \pm 0.2                     ^a$ &   15  -   150 & $ 1.22    ^{+ 0.09    }_{-    0.09    }\times10^{ 52  }$  &   17,18   \\
070810A &   2.17    & $ -0.75   ^{+ 0.83    }_{-    0.69    }$ &    ...                 & $ 42.23   ^{+ 6.62    }_{-    6.46    }$  &$  9.92    ^{+ 4.1 }_{-    2.88    }\times10^{ -8  }$      &   15  -   350 & $ 4.92    ^{+ 2.04    }_{-    1.43    }\times10^{ 51  }$  &   19  \\
071003  &   1.605   & $ -0.76   ^{+ 0.06    }_{-    0.07    }$ &    ...                 & $ 780 ^{+ 81  }_{-    70  }$  &$  1.22    ^{+ 0.19    }_{-    0.22    }\times10^{ -5  }$      &   20  -   4000    & $ 2.18    ^{+ 0.34    }_{-    0.39    }\times10^{ 53  }$  &   20  \\
071010B &   0.947   & $ -1.25   ^{+ 0.74    }_{-    0.49    }$ & $  -2.65   ^{+ 0.29    }_{-    0.49    }$ & $  52  ^{+ 10  }_{-    14  }$  &$  8.92    ^{+ 2.99    }_{-    5.99    }\times10^{ -7  }$      &   20  -   1000    & $ 6.47    ^{+ 2.17    }_{-    4.34    }\times10^{ 51  }$  &   21  \\
071020  &   2.145   & $  -0.65  ^{+ 0.27    }_{-    0.32    }$ &    ...                 & $ 322 ^{+ 80  }_{-    53  }$  &$  6.04    ^{+ 1.22    }_{-    3.88    }\times10^{ -6  }$      &   20  -   2000    & $ 2.25    ^{+ 0.45    }_{-    1.44    }\times10^{ 53  }$  &   22  \\
071117  &   1.331   & $  -1.53  ^{+ 0.15    }_{-    0.16    }$ &    ...                 & $ 278 ^{+ 236 }_{-    79  }$  &$  6.66    ^{+ 1.13    }_{-    2.95    }\times10^{ -6  }$      &   20  -   1000    & $ 9.95    ^{+ 1.69    }_{-    4.41    }\times10^{ 52  }$  &   23  \\
071227  &   0.383   & $ -0.7                    $ & ...                 & $ 1000                    $   &$( 3.5 \pm 1.1 )\times10^{ -6  }$              &   20  -   1300    & $ 2.52    ^{+ 0.79    }_{-    0.79    }\times10^{ 51  }$  &   24  \\
080207  &   2.0858  & $ -1.17   ^{+ 0.27    }_{-    0.27    }$ &    ...                 & $ 107.8   ^{+ 72.5    }_{-    72.5    }$  &$  1.0 \pm 0.3                     ^a$ &   15  -   150 & $ 4.22    ^{+ 1.27    }_{-    1.27    }\times10^{ 51  }$  &   25  \\
080319B &   0.937   & $ -0.86   ^{+ 0.01    }_{-    0.01    }$ & $  -3.59   ^{+ 0.45    }_{-    0.45    }$ & $  675 ^{+ 22  }_{-    22  }$  &$( 2.26    \pm 0.21    )\times10^{ -5  }$              &   20  -   7000    & $ 1.05    ^{+ 0.10    }_{-    0.10    }\times10^{ 53  }$  &   1   \\
080319C &   1.95    & $ -1.01   ^{+ 0.13    }_{-    0.13    }$ & $  -1.87   ^{+ 0.15    }_{-    0.63    }$ & $  307 ^{+ 141 }_{-    92  }$  &$  3.35    ^{+ 0.79    }_{-    0.7 }\times10^{ -6  }$      &   20  -   4000    & $ 9.46    ^{+ 2.23    }_{-    1.98    }\times10^{ 52  }$  &   26  \\
080411  &   1.03    & $ -1.51   ^{+ 0.04    }_{-    0.05    }$ &    ...                 & $ 259 ^{+ 35  }_{-    27  }$  &$( 1.28    \pm 0.16    )\times10^{ -5  }$              &   20  -   2000    & $ 9.33    ^{+ 1.17    }_{-    1.17    }\times10^{ 52  }$  &   27  \\
080413A &   2.433   & $ -1.2    ^{+ 0.1 }_{-    0.1 }$ &    ...                 & $ 170 ^{+ 80  }_{-    40  }$  &$  5.6 \pm 0.2                     ^a$ &   15  -   150 & $ 4.41    ^{+ 0.16    }_{-    0.16    }\times10^{ 52  }$  &   28,29   \\
080413B &   1.1 & $  -1.23  ^{+ 0.25    }_{-    0.25    }$ &    ...                 & $ 78  ^{+ 16  }_{-    16  }$  &$  1.4 \pm 0.2                     ^a$ &   15  -   150 & $ 1.55    ^{+ 0.06    }_{-    0.06    }\times10^{ 52  }$  &   1   \\
080603B &   2.69    & $  -1.20  ^{+ 0.26    }_{-    0.32    }$ &    ...                 & $ 200 ^{+ 131 }_{-    59  }$  &$  1.51    ^{+ 0.4 }_{-    0.38    }\times10^{ -6  }$      &   20  -   1000    & $ 1.11    ^{+ 0.29    }_{-    0.28    }\times10^{ 53  }$  &   30  \\
080605  &   1.6398  & $ -0.87   ^{+ 0.13    }_{-    0.12    }$ & $  -2.58   ^{+ 0.31    }_{-    0.84    }$ & $  297 ^{+ 46  }_{-    40  }$  &$( 1.6 \pm 0.33    )\times10^{ -5  }$              &   20  -   2000    & $ 3.33    ^{+ 0.69    }_{-    0.69    }\times10^{ 53  }$  &   31  \\
080607  &   3.036   & $ -0.76   ^{+ 0.07    }_{-    0.06    }$ & $  -2.57   ^{+ 0.18    }_{-    0.26    }$ & $  348 ^{+ 27  }_{-    27  }$  &$( 2.69    \pm 0.54    )\times10^{ -5  }$              &   20  -   4000    & $ 2.21    ^{+ 0.44    }_{-    0.44    }\times10^{ 54  }$  &   32  \\
080721  &   2.602   & $  -0.96  ^{+ 0.07    }_{-    0.07    }$ & $  -2.42   ^{+ 0.29    }_{-    0.29    }$ & $  497 ^{+ 62  }_{-    62  }$  &$( 2.11    \pm 0.35    )\times10^{ -5  }$              &   20  -   7000    & $ 1.11    ^{+ 0.18    }_{-    0.18    }\times10^{ 54  }$  &   1   \\
080804  &   2.2 & $ -0.88   ^{+ 0.1 }_{-    0.1 }$ &    ...                 & $ 315.1   ^{+ 67.4    }_{-    67.4    }$  &$( 7.3 \pm 0.88    )\times10^{ -7  }$              &   8   -   35000   & $ 2.86    ^{+ 0.34    }_{-    0.34    }\times10^{ 52  }$  &   33  \\
080810  &   3.35    & $ -1.2    ^{+ 0.1 }_{-    0.1 }$ &    -2.5                    & $ 580 ^{+ 850 }_{-    263 }$  &$  1.7 ^{+ 0.1 }_{-    0.2 }\times10^{ -5  }$      &   15  -   1000    & $ 2.39    ^{+ 0.14    }_{-    0.28    }\times10^{ 54  }$  &   34  \\
080913  &   6.7 & $ -0.82   ^{+ 0.75    }_{-    0.53    }$ &    -2.5                    & $ 121 ^{+ 232 }_{-    39  }$  &$( 1.4 \pm 0.058   )\times10^{ -6  }$              &   15  -   150 & $ 1.24    ^{+ 0.18    }_{-    0.18    }\times10^{ 53  }$  &   35,36   \\
080916A     &   0.689   & $  -0.99  ^{+ 0.05    }_{-    0.05    }$ &    ...                 & $ 208 ^{+ 11  }_{-    11  }$  &$( 4.87    \pm 0.27    )\times10^{ -7  }$              &   8   -   35000   & $ 1.08    ^{+ 0.06    }_{-    0.06    }\times10^{ 51  }$  &   1   \\
081007  &   0.5295  & $  -1.4   ^{+ 0.4 }_{-    0.4 }$ &    ...                 & $ 40  ^{+ 10  }_{-    10  }$  &$  2.2 \pm 0.2                     ^a$ &   25  -   900 & $ 4.35    ^{+ 0.40    }_{-    0.40    }\times10^{ 50  }$  &   1   \\
081008  &   1.9685  & $ -0.36   ^{+ 0.20    }_{-    0.20    }$ &    ...                 & $ 176.4   ^{+ 23.9    }_{-    23.9    }$  &$( 3.21    \pm 0.33    )\times10^{ -7  }$              &   8   -   35000   & $ 9.48    ^{+ 0.97    }_{-    0.97    }\times10^{ 51  }$  &   33  \\
081028  &   3.038   & $ 0.36    ^{+ 0.34    }_{-    0.34    }$ & $  -2.25   ^{+ 0.1 }_{-    0.1 }$ & $  59.66   ^{+ 5.91    }_{-    5.91    }$  &$( 7.04    \pm 0.65    )\times10^{ -7  }$              &   8   -   35000   & $ 4.91    ^{+ 0.45    }_{-    0.45    }\times10^{ 52  }$  &   33  \\
081118  &   2.58    & $ -0.68   ^{+ 0.09    }_{-    0.09    }$ &    ...                 & $ 98.99   ^{+ 5.01    }_{-    5.01    }$  &$( 6.73    \pm 0.23    )\times10^{ -7  }$              &   8   -   35000   & $ 3.99    ^{+ 0.14    }_{-    0.14    }\times10^{ 52  }$  &   33  \\
081121  &   2.512   & $ -0.21   ^{+ 0.28    }_{-    0.28    }$ & $  -1.86   ^{+ 0.09    }_{-    0.09    }$ & $  206.9   ^{+ 43.8    }_{-    43.8    }$  &$  5.16    ^{+ 1.53    }_{-    1.04    }\times10^{ -8  }$      &   8   -   35000   & $ 1.38    ^{+ 0.22    }_{-    0.22    }\times10^{ 53  }$  &   33  \\
081203A     &   2.1 & $  -1.29  ^{+ 0.15    }_{-    0.13    }$ &    ...                 & $ 497 ^{+ 244 }_{-    244 }$  &$  3.71    ^{+ 0.55    }_{-    0.48    }\times10^{ -7  }$      &   15  -   350 & $ 2.63    ^{+ 0.39    }_{-    0.34    }\times10^{ 52  }$  &   1   \\
081222  &   2.77    & $  -0.90  ^{+ 0.03    }_{-    0.03    }$ & $  -2.33   ^{+ 0.1 }_{-    0.1 }$ & $  167 ^{+ 8   }_{-    8   }$  &$( 1.76    \pm 0.058   )\times10^{ -6  }$              &   8   -   35000   & $ 1.01    ^{+ 0.03    }_{-    0.03    }\times10^{ 53  }$  &   1   \\
090102  &   1.547   & $  -0.97  ^{+ 0.01    }_{-    0.01    }$ &    ...                 & $ 461 ^{+ 15  }_{-    15  }$  &$( 2.93    \pm 0.091   )\times10^{ -6  }$              &   8   -   35000   & $ 4.79    ^{+ 0.15    }_{-    0.15    }\times10^{ 52  }$  &   1   \\
090424  &   0.544   & $  -1.02  ^{+ 0.01    }_{-    0.01    }$ & $  -3.26   ^{+ 0.18    }_{-    0.18    }$ & $  162 ^{+ 2.2 }_{-    2.2 }$  &$( 9.12    \pm 0.14    )\times10^{ -6  }$              &   8   -   35000   & $ 1.14    ^{+ 0.02    }_{-    0.02    }\times10^{ 52  }$  &   1   \\
090429B     &   9.4 & $ -0.69   ^{+ 0.91    }_{-    0.76    }$ &    ...                 & $ 46.21   ^{+ 10.66   }_{-    6.59    }$  &$  1.03    ^{+ 0.47    }_{-    0.32    }\times10^{ -7  }$      &   15  -   350 & $ 1.59    ^{+ 0.73    }_{-    0.49    }\times10^{ 53  }$  &   19  \\
090516  &   4.109   & $ -1.03   ^{+ 0.26    }_{-    0.31    }$ & $  -2.1    ^{+ 0.1 }_{-    0.2 }$ & $  51.4    ^{+ 20.1    }_{-    11.4    }$  &$  5.3 \pm 0.2                     ^a$ &   8   -   1000    & $ 8.70    ^{+ 0.33    }_{-    0.33    }\times10^{ 52  }$  &   37  \\
090519  &   3.85    & $ -0.58   ^{+ 0.22    }_{-    0.22    }$ &    ...                 & $ 120.5   ^{+ 13.8    }_{-    13.8    }$  &$( 2.25    \pm 0.17    )\times10^{ -7  }$              &   8   -   35000   & $ 3.46    ^{+ 0.26    }_{-    0.26    }\times10^{ 52  }$  &   33  \\
090529  &   2.625   & $ -0.75   ^{+ 0.03    }_{-    0.03    }$ &    ...                 & $ 199.9   ^{+ 6.74    }_{-    6.74    }$  &$( 3.006   \pm 0.063   )\times10^{ -6  }$              &   8   -   35000   & $ 1.82    ^{+ 0.04    }_{-    0.04    }\times10^{ 53  }$  &   33  \\
090618  &   0.54    & $ -0.91   ^{+ 0.03    }_{-    0.03    }$ & $  -2.42   ^{+ 0.07    }_{-    0.07    }$ & $  313.2   ^{+ 14.0    }_{-    14.0    }$  &$( 1.73    \pm 0.073   )\times10^{ -5  }$              &   8   -   35000   & $ 1.87    ^{+ 0.08    }_{-    0.08    }\times10^{ 52  }$  &   33  \\
090715B     &   3   & $  -1.1   ^{+ 0.37    }_{-    0.37    }$ &    ...                 & $ 134 ^{+ 41  }_{-    41  }$  &$( 9   \pm 2.5 )\times10^{ -7  }$              &   20  -   2000    & $ 8.78    ^{+ 2.44    }_{-    2.44    }\times10^{ 52  }$  &   1   \\
090809  &   2.737   & $ -0.47   ^{+ 0.05    }_{-    0.05    }$ & $  -2.16   ^{+ 0.07    }_{-    0.07    }$ & $  193.4   ^{+ 11.2    }_{-    11.2    }$  &$( 7.231   \pm 0.6 )\times10^{ -6  }$              &   8   -   35000   & $ 3.40    ^{+ 0.28    }_{-    0.28    }\times10^{ 53  }$  &   33  \\
090812  &   2.452   & $  -1.03  ^{+ 0.07    }_{-    0.07    }$ &    ...                 & $ 586 ^{+ 192 }_{-    192 }$  &$  2.77    \pm 0.28                        ^a$ &   100 -   1000    & $ 1.02    ^{+ 0.10    }_{-    0.10    }\times10^{ 53  }$  &   1   \\
090926B     &   1.24    & $  -0.19  ^{+ 0.06    }_{-    0.06    }$ &    ...                 & $ 95.6    ^{+ 1.9 }_{-    1.9 }$  &$( 4.73    \pm 0.28    )\times10^{ -7  }$              &   8   -   35000   & $ 4.46    ^{+ 0.26    }_{-    0.26    }\times10^{ 51  }$  &   1   \\
090927  &   1.37    & $ -0.68   ^{+ 0.05    }_{-    0.05    }$ & $  -2.12   ^{+ 0.01    }_{-    0.01    }$ & $  59.67   ^{+ 1.81    }_{-    1.81    }$  &$( 9.379   \pm 0.23    )\times10^{ -6  }$              &   8   -   35000   & $ 9.30    ^{+ 0.23    }_{-    0.23    }\times10^{ 52  }$  &   33  \\
091018  &   0.971   & $  -1.53  ^{+ 0.48    }_{-    0.48    }$ &    ...                 & $ 28  ^{+ 13  }_{-    13  }$  &$( 4.32    \pm 0.95    )\times10^{ -7  }$              &   20  -   1000    & $ 4.90    ^{+ 1.08    }_{-    1.08    }\times10^{ 51  }$  &   1   \\
091020  &   1.71    & $ -1.20   ^{+ 0.06    }_{-    0.06    }$ & $  -2.29   ^{+ 0.18    }_{-    0.18    }$ & $  187 ^{+ 25  }_{-    25  }$  &$( 1.88    \pm 0.026   )\times10^{ -6  }$              &   8   -   35000   & $ 3.44    ^{+ 0.04    }_{-    0.04    }\times10^{ 52  }$  &   1   \\
091024  &   1.092   & $ -1.5    ^{+ 0.4 }_{-    0.4 }$ &    ...                 & $ 280 ^{+ 120 }_{-    120 }$  &$  3.46    ^{+ 0.53    }_{-    0.46    }\times10^{ -7  }$      &   15  -   350 & $ 4.08    ^{+ 0.62    }_{-    0.54    }\times10^{ 51  }$  &   38  \\
091029  &   2.752   & $ -1.46   ^{+ 0.27    }_{-    0.27    }$ &    ...                 & $ 61.4    ^{+ 17.5    }_{-    17.5    }$  &$  1.8 \pm 0.1                     ^a$ &   15  -   150 & $ 1.40    ^{+ 0.08    }_{-    0.08    }\times10^{ 52  }$  &   39  \\
091127  &   0.49    & $ -0.68   ^{+ 0.05    }_{-    0.05    }$ & $  -2.12   ^{+ 0.01    }_{-    0.01    }$ & $  59.67   ^{+ 1.81    }_{-    1.81    }$  &$( 9.379   \pm 0.23    )\times10^{ -6  }$              &   8   -   35000   & $ 7.71    ^{+ 0.19    }_{-    0.19    }\times10^{ 51  }$  &   33  \\
091208B     &   1.063   & $  -1.29  ^{+ 0.04    }_{-    0.04    }$ &    ...                 & $ 119 ^{+ 7   }_{-    7   }$  &$( 2.56    \pm 0.097   )\times10^{ -6  }$              &   8   -   35000   & $ 1.81    ^{+ 0.06    }_{-    0.06    }\times10^{ 52  }$  &   1   \\
100425A     &   1.755   & $ -0.53   ^{+ 2.83    }_{-    1.46    }$ &    ...                 & $ <36.02                  $   &$  4.74    ^{+ 3.46    }_{-    1.97    }\times10^{ -8  }$      &   15  -   350 & $ <1.41                   \times10^{  51  }$  &   40  \\
100615A     &   1.398   & $ -1.24   ^{+ 0.08    }_{-    0.06    }$ & $  -2.27   ^{+ 0.11    }_{-    0.12    }$ & $  85.73   ^{+ 7.82    }_{-    9.33    }$  &$  8.3 \pm 0.2                     ^a$ &   8   -   1000    & $ 1.06    ^{+ 0.03    }_{-    0.03    }\times10^{ 52  }$  &   41  \\
100621A     &   0.542   & $  -1.70  ^{+ 0.13    }_{-    0.13    }$ & $  -2.45   ^{+ 0.15    }_{-    0.15    }$ & $  95  ^{+ 15  }_{-    15  }$  &$( 1.7 \pm 0.13    )\times10^{ -6  }$              &   20  -   2000    & $ 3.24    ^{+ 0.25    }_{-    0.25    }\times10^{ 51  }$  &   1   \\
100728A     &   1.567   & $ -0.47   ^{+ 0.15    }_{-    0.15    }$ & $  -2.5    ^{+ 0.2 }_{-    0.3 }$ & $  390 ^{+ 27  }_{-    25  }$  &$( 4.2 \pm 0.7 )\times10^{ -6  }$              &   20  -   10000   & $ 6.45    ^{+ 1.08    }_{-    1.08    }\times10^{ 52  }$  &   42  \\
100728B     &   2.106   & $  -0.90  ^{+ 0.07    }_{-    0.07    }$ &    ...                 & $ 130 ^{+ 9   }_{-    9   }$  &$( 5.43    \pm 0.35    )\times10^{ -7  }$              &   8   -   35000   & $ 1.97    ^{+ 0.13    }_{-    0.13    }\times10^{ 52  }$  &   1   \\
100814A     &   1.44    & $ -0.55   ^{+ 0.3 }_{-    0.3 }$ &    ...                 & $ 147 ^{+ 12  }_{-    10  }$  &$( 7.5 \pm 2.5 )\times10^{ -7  }$              &   20  -   2000    & $ 1.08    ^{+ 0.36    }_{-    0.36    }\times10^{ 52  }$  &   43  \\
100816A     &   0.8049  & $ -0.31   ^{+ 0.05    }_{-    0.05    }$ & $  -2.77   ^{+ 0.17    }_{-    0.17    }$ & $  136.7   ^{+ 4.73    }_{-    4.73    }$  &$  15.59   \pm 0.25                        ^a$ &   10  -   1000    & $ 7.38    ^{+ 0.12    }_{-    0.12    }\times10^{ 51  }$  &   44,45   \\
100906A     &   1.727   & $ -1.1    ^{+ 0.1 }_{-    0.1 }$ & $  -2.2    ^{+ 0.2 }_{-    0.3 }$ & $  180 ^{+ 45  }_{-    40  }$  &$( 2.7 \pm 0.3 )\times10^{ -6  }$              &   20  -   2000    & $ 6.90    ^{+ 0.77    }_{-    0.77    }\times10^{ 52  }$  &   46  \\
101213A     &   0.414   & $ -1.1    ^{+ 0.07    }_{-    0.07    }$ & $  -2.35   ^{+ 0.29    }_{-    0.72    }$ & $  309.7   ^{+ 48.9    }_{-    40.0    }$  &$  4.67    \pm 0.32                        ^a$ &   10  -   1000    & $ 6.32    ^{+ 0.43    }_{-    0.43    }\times10^{ 50  }$  &   47  \\
101219B     &   0.55    & $ -0.33   ^{+ 0.36    }_{-    0.36    }$ & $  -2.12   ^{+ 0.12    }_{-    0.12    }$ & $  70  ^{+ 8   }_{-    8   }$  &$  2.0 \pm 0.2                     ^a$ &   10  -   1000    & $ 3.81    ^{+ 0.38    }_{-    0.38    }\times10^{ 50  }$  &   48  \\
110205A     &   2.22    & $  -1.52  ^{+ 0.14    }_{-    0.14    }$ &    ...                 & $ 222 ^{+ 74  }_{-    74  }$  &$( 5.1 \pm 0.7 )\times10^{ -7  }$              &   20  -   1200    & $ 2.65    ^{+ 0.36    }_{-    0.36    }\times10^{ 52  }$  &   1   \\
110213A     &   1.46    & $ -1.44   ^{+ 0.05    }_{-    0.05    }$ &    ...                 & $ 98.4    ^{+ 8.5 }_{-    6.9 }$  &$  17.7    \pm 0.5                     ^a$ &   10  -   1000    & $ 2.23    ^{+ 0.06    }_{-    0.06    }\times10^{ 52  }$  &   49  \\
110422A     &   1.77    & $ -0.53   ^{+ 0.17    }_{-    0.14    }$ & $  -2.65   ^{+ 0.28    }_{-    0.62    }$ & $  246 ^{+ 37  }_{-    34  }$  &$( 1.2 \pm 0.15    )\times10^{ -5  }$              &   20  -   2000    & $ 2.90    ^{+ 0.36    }_{-    0.36    }\times10^{ 51  }$  &   50  \\
110503A     &   1.613   & $  -0.98  ^{+ 0.08    }_{-    0.08    }$ & $  -2.7    ^{+ 0.3 }_{-    0.3 }$ & $  219 ^{+ 19  }_{-    19  }$  &$( 10  \pm 1   )\times10^{ -6  }$              &   20  -   5000    & $ 1.89    ^{+ 0.19    }_{-    0.19    }\times10^{ 53  }$  &   1   \\
110715A     &   0.82    & $ -1.23   ^{+ 0.09    }_{-    0.08    }$ & $  -2.7    ^{+ 0.2 }_{-    0.5 }$ & $  120 ^{+ 12  }_{-    11  }$  &$( 1.1 \pm 0.1 )\times10^{ -5  }$              &   20  -   10000   & $ 4.31    ^{+ 0.39    }_{-    0.39    }\times10^{ 52  }$  &   51  \\
110731A     &   2.83    & $ -0.8    ^{+ 0.03    }_{-    0.03    }$ & $  -2.98   ^{+ 0.3 }_{-    0.3 }$ & $  304 ^{+ 13  }_{-    13  }$  &$  20.9    \pm 0.5                     ^a$ &   10  -   1000    & $ 3.06    ^{+ 0.07    }_{-    0.07    }\times10^{ 53  }$  &   52  \\
110801A     &   1.858   & $ -1.7    ^{+ 0.12    }_{-    0.15    }$ &    -2.5                    & $ 140 ^{+ 1270    }_{-    50  }$  &$  8.91    ^{+ 1.68    }_{-    1.4 }\times10^{ -8  }$      &   15  -   350 & $ 4.43    ^{+ 0.84    }_{-    0.70    }\times10^{ 51  }$  &   53  \\
110808A     &   1.348   & $ -1.07   ^{+ 0.12    }_{-    0.11    }$ &    ...                 & $ 4238    ^{+ 3270    }_{-    1530    }$  &$( 1.1 \pm 0.2 )\times10^{ -4  }$              &   20  -   10000   & $ 8.96    ^{+ 1.63    }_{-    1.63    }\times10^{ 53  }$  &   54  \\
110818A     &   3.36    & $ -1.33   ^{+ 0.08    }_{-    0.08    }$ &    ...                 & $ 256.3   ^{+ 55.3    }_{-    55.3    }$  &$  5.0 \pm 1.4                     ^a$ &   10  -   1000    & $ 7.23    ^{+ 2.03    }_{-    2.03    }\times10^{ 52  }$  &   55  \\
111008A     &   4.9898  & $ -1.36   ^{+ 0.24    }_{-    0.21    }$ &    ...                 & $ 149 ^{+ 52  }_{-    28  }$  &$( 1.4 \pm 0.3 )\times10^{ -6  }$              &   20  -   2000    & $ 4.95    ^{+ 1.06    }_{-    1.06    }\times10^{ 53  }$  &   56  \\
111107A     &   2.893   & $ -1.38   ^{+ 0.21    }_{-    0.21    }$ &    ...                 & $ 108 ^{+ 32  }_{-    32  }$  &$  2.6 \pm 0.3                     ^a$ &   10  -   1000    & $ 1.81    ^{+ 0.21    }_{-    0.21    }\times10^{ 52  }$  &   57  \\
111123A     &   3.1516  & $ -1.30   ^{+ 0.26    }_{-    0.24    }$ &    ...                 & $ 107.79  ^{+ 125.38  }_{-    25.03   }$  &$  7.89    ^{+ 1.1 }_{-    0.89    }\times10^{ -8  }$      &   15  -   350 & $ 9.80    ^{+ 1.37    }_{-    1.11    }\times10^{ 51  }$  &   40  \\
111228A     &   0.714   & $ -1.9    ^{+ 0.1 }_{-    0.1 }$ & $  -2.7    ^{+ 0.3 }_{-    0.3 }$ & $  34  ^{+ 3   }_{-    3   }$  &$  27  \pm 1                       ^a$ &   10  -   1000    & $ 6.67    ^{+ 0.25    }_{-    0.25    }\times10^{ 51  }$  &   58  \\
120119A     &   1.728   & $ -0.98   ^{+ 0.03    }_{-    0.03    }$ & $  -2.36   ^{+ 0.09    }_{-    0.09    }$ & $  189.2   ^{+ 8.3 }_{-    8.3 }$  &$  16.86   \pm 0.39                        ^a$ &   10  -   1000    & $ 5.98    ^{+ 0.14    }_{-    0.14    }\times10^{ 52  }$  &   59  \\
120326A     &   1.798   & $ -0.98   ^{+ 0.14    }_{-    0.14    }$ & $  -2.53   ^{+ 0.15    }_{-    0.15    }$ & $  46.45   ^{+ 3.67    }_{-    3.67    }$  &$  3.1 \pm 0.05                        ^a$ &   10  -   1000    & $ 5.91    ^{+ 0.10    }_{-    0.10    }\times10^{ 51  }$  &   60  \\
120327A     &   2.813   & $ -1.14   ^{+ 0.26    }_{-    0.28    }$ &    ...                 & $ 106.09  ^{+ 80.1    }_{-    22.76   }$  &$  3.88    ^{+ 0.65    }_{-    0.52    }\times10^{ -7  }$      &   15  -   350 & $ 3.42    ^{+ 0.57    }_{-    0.46    }\times10^{ 52  }$  &   40  \\
120712A     &   4.1745  & $ -0.6    ^{+ 0.2 }_{-    0.2 }$ & $  -1.8    ^{+ 0.2 }_{-    0.2 }$ & $  124 ^{+ 26  }_{-    26  }$  &$  3.5 \pm 0.2                     ^a$ &   10  -   1000    & $ 1.35    ^{+ 0.08    }_{-    0.08    }\times10^{ 53  }$  &   61  \\
120714B     &   0.3984  & $ -0.29   ^{+ 0.96    }_{-    0.8 }$ &    ...                 & $ 60.8    ^{+ 25.92   }_{-    10.22   }$  &$  1.8 ^{+ 1.13    }_{-    0.68    }\times10^{ -8  }$      &   15  -   350 & $ 1.15    ^{+ 0.72    }_{-    0.44    }\times10^{ 49  }$  &   40  \\
120724A     &   1.48    & $ -0.75   ^{+ 2.34    }_{-    1.23    }$ &    ...                 & $ <31.9                   $   &$  1.95    ^{+ 1.58    }_{-    0.87    }\times10^{ -8  }$      &   15  -   350 & $ <4.19                   \times10^{  50  }$  &   40  \\
120802A     &   3.796   & $ -0.96   ^{+ 0.60    }_{-    0.53    }$ &    ...                 & $ 52.96   ^{+ 12.58   }_{-    6.84    }$  &$  1.85    ^{+ 0.34    }_{-    0.27    }\times10^{ -7  }$      &   15  -   350 & $ 3.51    ^{+ 0.65    }_{-    0.51    }\times10^{ 52  }$  &   40  \\
120811C     &   2.671   & $ -1.19   ^{+ 0.32    }_{-    0.30    }$ &    ...                 & $ 46.26   ^{+ 4.32    }_{-    4.14    }$  &$  2.47    ^{+ 0.25    }_{-    0.22    }\times10^{ -7  }$      &   15  -   350 & $ 2.23    ^{+ 0.23    }_{-    0.20    }\times10^{ 52  }$  &   40  \\
120907A     &   0.97    & $ -0.75   ^{+ 0.25    }_{-    0.25    }$ &    ...                 & $ 154.5   ^{+ 32.9    }_{-    32.9    }$  &$  4.3 \pm 0.4                     ^a$ &   10  -   1000    & $ 2.53    ^{+ 0.24    }_{-    0.24    }\times10^{ 51  }$  &   62  \\
120909A     &   3.93    & $ -1.3    ^{+ 0.1 }_{-    0.1 }$ &    ...                 & $ 370 ^{+ 140 }_{-    140 }$  &$  3.0 \pm 0.2                     ^a$ &   10  -   1000    & $ 7.65    ^{+ 0.51    }_{-    0.51    }\times10^{ 52  }$  &   63  \\
120922A     &   3.1 & $ -1.6    ^{+ 0.7 }_{-    0.7 }$ & $  -2.3    ^{+ 0.1 }_{-    0.1 }$ & $  37.7    ^{+ 3.5 }_{-    3.5 }$  &$  3.4 \pm 0.3                     ^a$ &   10  -   1000    & $ 3.02    ^{+ 0.27    }_{-    0.27    }\times10^{ 52  }$  &   64  \\
121027A     &   1.77    & $ -1.49   ^{+ 0.43    }_{-    0.39    }$ &    ...                 & $ 61.75   ^{+ 437.65  }_{-    13.25   }$  &$  8.34    ^{+ 2.24    }_{-    1.64    }\times10^{ -8  }$      &   15  -   350 & $ 2.92    ^{+ 0.79    }_{-    0.57    }\times10^{ 53  }$  &   40  \\
121128A     &   2.2 & $ -0.80   ^{+ 0.12    }_{-    0.12    }$ & $  -2.41   ^{+ 0.1 }_{-    0.1 }$ & $  62.2    ^{+ 4.6 }_{-    4.6 }$  &$  17.9    \pm 0.5                     ^a$ &   10  -   1000    & $ 6.67    ^{+ 0.19    }_{-    0.19    }\times10^{ 52  }$  &   65  \\
121211A     &   1.023   & $ -0.3    ^{+ 0.34    }_{-    0.34    }$ &    ...                 & $ 95.96   ^{+ 12.6    }_{-    12.6    }$  &$  2.402   \pm 0.202                       ^a$ &   10  -   1000    & $ 1.25    ^{+ 0.11    }_{-    0.11    }\times10^{ 51  }$  &   66  \\
130215A     &   0.597   & $ -1  ^{+ 0.2 }_{-    0.2 }$ & $  -1.6    ^{+ 0.03    }_{-    0.03    }$ & $  155 ^{+ 63  }_{-    63  }$  &$  3.5 \pm 0.3                     ^a$ &   10  -   1000    & $ 2.16    ^{+ 0.18    }_{-    0.18    }\times10^{ 51  }$  &   67  \\
130408A     &   3.758   & $ -0.7    ^{+ 0.15    }_{-    0.15    }$ & $  -2.3    ^{+ 0.3 }_{-    0.3 }$ & $  272 ^{+ 40  }_{-    40  }$  &$( 5.2 \pm 0.5 )\times10^{ -6  }$              &   20  -   10000   & $ 6.12    ^{+ 0.59    }_{-    0.59    }\times10^{ 53  }$  &   68  \\
130420A     &   1.297   & $ -1  ^{+ 0.13    }_{-    0.13    }$ &    ...                 & $ 56  ^{+ 3   }_{-    3   }$  &$  5.2 \pm 0.4                     ^a$ &   10  -   1000    & $ 3.65    ^{+ 0.28    }_{-    0.28    }\times10^{ 51  }$  &   69  \\
130427A     &   0.3399  & $ -0.789  ^{+ 0.003   }_{-    0.003   }$ & $  -3.06   ^{+ 0.02    }_{-    0.02    }$ & $  830 ^{+ 5   }_{-    5   }$  &$  1052    \pm 2                       ^a$ &   8   -   1000    & $ 1.90    ^{+ 0.00    }_{-    0.00    }\times10^{ 53  }$  &   70  \\
130505A     &   2.27    & $ -0.31   ^{+ 0.09    }_{-    0.09    }$ & $  -2.26   ^{+ 0.07    }_{-    0.07    }$ & $  604 ^{+ 49  }_{-    49  }$  &$( 6.9 \pm 0.3 )\times10^{ -5  }$              &   20  -   1200    & $ 3.98    ^{+ 0.17    }_{-    0.17    }\times10^{ 54  }$  &   71  \\
130514A     &   3.6 & $ -1.44   ^{+ 0.17    }_{-    0.15    }$ &    ...                 & $ 110 ^{+ 42  }_{-    21  }$  &$  2.06    ^{+ 0.23    }_{-    0.18    }\times10^{ -7  }$      &   15  -   350 & $ 3.82    ^{+ 0.43    }_{-    0.33    }\times10^{ 52  }$  &   72  \\
130606A     &   5.913   & $ -1.14   ^{+ 0.15    }_{-    0.15    }$ &    ...                 & $ 294 ^{+ 90  }_{-    50  }$  &$  3.15    ^{+ 0.56    }_{-    0.46    }\times10^{ -7  }$      &   15  -   350 & $ 2.04    ^{+ 0.36    }_{-    0.30    }\times10^{ 53  }$  &   73  \\
130610A     &   2.092   & $ -1  ^{+ 0.1 }_{-    0.1 }$ &    ...                 & $ 294.9   ^{+ 42.9    }_{-    42.9    }$  &$  4.5 \pm 0.9                     ^a$ &   10  -   1000    & $ 2.55    ^{+ 0.51    }_{-    0.51    }\times10^{ 52  }$  &   74  \\
130612A     &   2.006   & $ -1.3    ^{+ 0.3 }_{-    0.3 }$ &    ...                 & $ 61.9    ^{+ 10.5    }_{-    10.5    }$  &$  4.1 \pm 0.2                     ^a$ &   10  -   1000    & $ 9.48    ^{+ 0.46    }_{-    0.46    }\times10^{ 51  }$  &   75  \\
130701A     &   1.155   & $ -1.1    ^{+ 0.1 }_{-    0.1 }$ &    ...                 & $ 89  ^{+ 4   }_{-    4   }$  &$( 4.3 \pm 0.4 )\times10^{ -6  }$              &   20  -   10000   & $ 4.27    ^{+ 0.40    }_{-    0.40    }\times10^{ 52  }$  &   76  \\
130831A &   0.4791  & $ -1.51   ^{+ 0.1 }_{-    0.1 }$ & $  -2.8    ^{+ 0.1 }_{-    0.1 }$ & $  67  ^{+ 4   }_{-    4   }$  &$( 2.6 \pm 0.3 )\times10^{ -6  }$              &   20  -   10000   & $ 3.42    ^{+ 0.39    }_{-    0.39    }\times10^{ 51  }$  &   77  \\
130907A     &   1.238   & $ -0.65   ^{+ 0.03    }_{-    0.03    }$ & $  -2.22   ^{+ 0.05    }_{-    0.05    }$ & $  390 ^{+ 16  }_{-    16  }$  &$( 2.2 \pm 0.1 )\times10^{ -5  }$              &   20  -   10000   & $ 1.82    ^{+ 0.08    }_{-    0.08    }\times10^{ 53  }$  &   78  \\
131030A     &   1.295   & $ -0.71   ^{+ 0.12    }_{-    0.12    }$ & $  -2.95   ^{+ 0.28    }_{-    0.28    }$ & $  177 ^{+ 10  }_{-    10  }$  &$( 10  \pm 1   )\times10^{ -6  }$              &   20  -   10000   & $ 1.08    ^{+ 0.11    }_{-    0.11    }\times10^{ 53  }$  &   79  \\

\enddata
\\
\begin{flushleft}
{Notes: $^a$ For these GRBs, the peak flux is in units of $\rm photons~cm^{-2}~s^{-1}.$ \\
$^b$ For those GRBs with $\beta$ value, the spectra are described well by Band model. But for the GRBs without $\beta$ value, the spectra are described by
power-low with an exponential cutoff model. \\
Reference:[1]\cite{Nava2012}, [2]\cite{Golenetskii2005a}, [3]\cite{Golenetskii2005b}, [4]\cite{Sugita2009}, [5]\cite{Butler2007}, [6]\cite{Golenetskii2005c}, [7]\cite{Barbier2006}, [8]\cite{Golenetskii2006a}, [9]\cite{Sakamoto2006}, [10]\cite{Golenetskii2006b}, [11]\cite{Stamatikos2006}, [12]\cite{Golenetskii2006c}, [13]\cite{Golenetskii2006d}, [14]\cite{Golenetskii2006e}, [15]\cite{Golenetskii2006f}, [16]\cite{Golenetskii2007a}, [17]\cite{Ohno2007}, [18]\cite{Barbier2007}, [19]\cite{Butler2010}, [20]\cite{Golenetskii2007b}, [21]\cite{Golenetskii2007c}, [22]\cite{Golenetskii2007d}, [23]\cite{Golenetskii2007e}, [24]\cite{Golenetskii2007f}, [25]\cite{Stamatikos2008}, [26]\cite{Golenetskii2008a}, [27]\cite{Golenetskii2008b}, [28]\cite{Tueller2008}, [29]\cite{Ohno2008}, [30]\cite{Golenetskii2008c}, [31]\cite{Golenetskii2008d}, [32]\cite{Golenetskii2008e}, [33]\cite{Nava2011}, [34]\cite{Sakamoto2008}, [35]\cite{Pal'Shin2008}, [36]\cite{Greiner2009}, [37]\cite{McBreen2009}, [38]\cite{Golenetskii2009}, [39]\cite{Barthelmy2009},
[40]Butler's
website\footnote{http://butler.lab.asu.edu/Swift/bat\_spec\_table.html}
\& \emph{Swift} official
website\footnote{http://Swift.gsfc.nasa.gov/archive/grb\_table/}.
[41]\cite{Foley2010}, [42]\cite{Golenetskii2010a}, [43]\cite{Golenetskii2010b}, [44]\cite{Fitzpatrick2010a}, [45]\cite{Fitzpatrick2010b}, [46]\cite{Golenetskii2010c}, [47]\cite{Gruber2010}, [48]\cite{Horst2010}, [49]\cite{Foley2011}, [50]\cite{Golenetskii2011a}, [51]\cite{Golenetskii2011b}, [52]\cite{Golenetskii2011c}, [53]\cite{Sakamoto2011}, [54]\cite{Golenetskii2011d}, [55]\cite{Xiong2011}, [56]\cite{Golenetskii2011e}, [57]\cite{Pelassa2011}, [58]\cite{Briggs2011}, [59]\cite{Gruber2012a}, [60]\cite{Collazzi2012}, [61]\cite{Gruber2012b}, [62]\cite{Younes2012}, [63]\cite{Chaplin2012}, [64]\cite{Younes2012a}, [65]\cite{McGlynn2012}, [66]\cite{Yu2012}, [67]\cite{Younes2013}, [68]\cite{Golenetskii2013a}, [69]\cite{Xiong2013}, [70]\cite{Kienlin2013}, [71]\cite{Golenetskii2013b}, [72]\cite{Pal'Shin2013}, [73]\cite{Barthelmy2013}, [74]\cite{Fitzpatrick2013}, [75]\cite{Fitzpatrick2013a}, [76]\cite{Golenetskii2013c}, [77]\cite{Golenetskii2013d}, [78]\cite{Golenetskii2013e}, [79]\cite{Golenetskii2013f}.}
\end{flushleft}
\end{deluxetable}

\begin{thebibliography}{}

\bibitem[\protect\citeauthoryear{Band et al.}{1993}]{Band93}Band, D., Matteson, J., Ford, L., et al., 1993, ApJ, 413, 281

\bibitem[Barbier et al.(2006)]{Barbier2006}Barbier, L., Barthelmy, S., Cummings, J., et al., 2006, GCN, 4518

\bibitem[Barbier et al.(2007)]{Barbier2007}Barbier, L., Barthelmy, S., Cummings, J., et al., 2007, GCN, 6623

\bibitem[Barger et al.(2000)]{Barger2000}Barger, A. J., Cowie, L. L., \& Richards, E. A., 2000, AJ, 119, 2092

\bibitem[Barthelmy et al.(2009)]{Barthelmy2009}Barthelmy, S. D., Baumgartner, W. H., Cummings, J. R., et al., 2009, GCN, 10103

\bibitem[Barthelmy et al.(2013)]{Barthelmy2013}Barthelmy, S. D., Baumgartner, W. H., Cummings, J. R., 2013, GCN, 10103

\bibitem[\protect\citeauthoryear{Bloom et al.}{2008}]{Bloom08}Bloom, J. S., Butler, N. R., \& Perley, D. A. 2008, in AIP Conf.
Ser. 1000, Gamma-Ray Bursts 2007: Proceedings of the Santa Fe Conf.,
ed.M. Galassi, D. Palmer, \& E. Fenimore (Melville, NY: AIP), 11

\bibitem[\protect\citeauthoryear{Bloom et al.}{2001}]{Bloom01}Bloom, J. S., Frail, D. A., \& Sari, R., 2001, AJ, 112, 2879

\bibitem[Bouwens et al.(2011)]{Bouwens2011}Bouwens, R. J., Illingworth, G. D., Labbe, I., et al. 2011, Nature, 479, 504

\bibitem[Briggs \& Younes(2011)]{Briggs2011}Briggs, M. S. \& Younes, G., 2011, GCN, 12744

\bibitem[Bromberg et al.(2013)]{Bromberg13}Bromberg, O., Nakar, E., Piran, T., \& Sari,
R., 2013, ApJ, 764, 179

\bibitem[Bromm \& Loeb(2002)]{BrommLoeb2002} Broom, V., \& Loeb, A., 2002, ApJ, 575, 111

\bibitem[\protect\citeauthoryear{Bromm \& Loeb}{2012}]{Bromm12}Bromm, V., Loeb, A., 2012, in Gamma-ray Bursts, ed. C. Kouveliotou, S. E. Woosley, R. A. M. J. Wijers (Cambridge: Cambridge Univ.
Press), arXiv:0706.2445v2

\bibitem[Butler et al.(2007)]{Butler2007} Butler, N. R., Kocevski, D., Bloom, J. S., \& Curtis, J. L., 2007, ApJ, 671, 656

\bibitem[Butler et al.(2010)]{Butler2010} Butler, N. R., Bloom, J. S. \& Poznanski, D. 2010, ApJ, 711, 495

\bibitem[\protect\citeauthoryear{Cao et al.}{2011}]{Cao11}Cao, X. F., Yu, Y. W., Cheng, K. S., \& Zheng, X. P., 2011, MNRAS, 416, 2174

\bibitem[\protect\citeauthoryear{Castro-Tirado et al.}{2013}]{CastroTirado14}Castro-Tirado, A. J., S\'{a}nchez-Ram\'{\i}rez, R., Ellison, S. L., et al., 2013, arXiv:1312.5631

\bibitem[Chaplin(2012)]{Chaplin2012}Chaplin, V., 2012, GCN, 13737

\bibitem[\protect\citeauthoryear{Cheng et al.}{2010}]{Cheng10}Cheng, K. S., Yu, Y., \& Harko, T., 2010, Phys. Rev. Lett., 104, 241102

\bibitem[\protect\citeauthoryear{Cobb et al.}{2006}]{Cobb06}Cobb, B. E., Bailyn, C. D., van Dokkum, P. G., \& Natarajan, P. 2006,
ApJ, 645, L113

\bibitem[Collazzi(2012)]{Collazzi2012}Collazzi, A. C., 2012, GCN, 13145

\bibitem[\protect\citeauthoryear{Coward}{2007}]{Coward07}Coward, D., 2007, New Astronomy Reviews, 51, 539

\bibitem[\protect\citeauthoryear{Dai et al.}{2004}]{Dai04} Dai, Z. G., Liang, E. W., \& Xu, D., 2004, ApJ, 612, L101

\bibitem[Efron \& Petrosian (1992)]{EfronPetrosian1992}Efron, B. \& Petrosian, V. 1992, ApJ, 399, 345

\bibitem[\protect\citeauthoryear{Elliott et al.}{2012}]{Elliott12}Elliott, J., Greiner, J., Khochfar, S., et al. 2012, A\&A, 539, A113

\bibitem[\protect\citeauthoryear{Elliott et al.}{2014}]{Elliott14}Elliott, J., Khochfar, S., Greiner, J., Dalla Vecchia, C., 2014, arXiv:1408.2526

\bibitem[Fenimore \& Ramire Ruiz (2000)]{FenimoreRamireRuiz2000}Fenimore, E. E. \& Ramirez-Ruiz, E. 2000, arXiv:astro-ph/0004176

\bibitem[Firmani et al.(2004)]{Firmani2004}Firmani, C., Avila-Reese, V., Ghisellini, G., \& Tutukov, A. V. 2004, ApJ, 611, 1033

\bibitem[Fitzpatrick(2010a)]{Fitzpatrick2010a}Fitzpatrick, G., 2010a, GCN, 11124

\bibitem[Fitzpatrick(2010b)]{Fitzpatrick2010b}Fitzpatrick, G., 2010b, GCN, 11128

\bibitem[Fitzpatrick \& Pelassa(2013)]{Fitzpatrick2013}Fitzpatrick, G. \& Pelassa, V., 2013, GCN, 14858

\bibitem[Fitzpatrick(2013)]{Fitzpatrick2013a}Fitzpatrick, G., 2013, GCN, 14896

\bibitem[Foley \& Briggs(2010)]{Foley2010}Foley, S. \& Briggs, M., 2010, GCN, 10851

\bibitem[Foley(2011)]{Foley2011}Foley, S., 2011, GCN, 10851

\bibitem[Fynbo et al.(2006)]{Fynbo06}Fynbo, J. P. U., Watson, D., Thone, C. C., et al., 2006, Natur, 444, 1047

\bibitem[Gal-Yam et al. (2006)]{Gal-Yam06}Gal-Yam, A., Fox, D. B., Price, P. A., et al., 2006, Natur, 444, 1053

\bibitem[\protect\citeauthoryear{Gehrels et al.}{2006}]{Gehrels06} Gehrels, N., Norris, J. P., Barthelmy, S. D., et al., 2006, Natur, 444, 1044

\bibitem[\protect\citeauthoryear{Gehrels et al.}{2009}]{Gehrels09} Gehrels, N., Ramirez-Ruiz, E., \& Fox, D. B., 2009, ARA\&A, 47, 567

\bibitem[Golenetskii et al.(2005a)]{Golenetskii2005a}Golenetskii, S., Aptekar, R., Mazets, E., et al., 2005a, GCN, 3179

\bibitem[Golenetskii et al.(2005b)]{Golenetskii2005b}Golenetskii, S., Aptekar, R., Mazets, E., et al., 2005b, GCN, 3518

\bibitem[Golenetskii et al.(2005c)]{Golenetskii2005c}Golenetskii, S., Aptekar, R., Mazets, E., et al., 2005c, GCN, 4238

\bibitem[Golenetskii et al.(2006a)]{Golenetskii2006a}Golenetskii, S., Aptekar, R., Mazets, E., et al., 2006a, GCN, 4599

\bibitem[Golenetskii et al.(2006b)]{Golenetskii2006b}Golenetskii, S., Aptekar, R., Mazets, E., et al., 2006b, GCN, 5264

\bibitem[Golenetskii et al.(2006c)]{Golenetskii2006c}Golenetskii, S., Aptekar, R., Mazets, E., et al., 2006c, GCN, 5459

\bibitem[Golenetskii et al.(2006d)]{Golenetskii2006d}Golenetskii, S., Aptekar, R., Mazets, E., et al., 2006d, GCN, 5722

\bibitem[Golenetskii et al.(2006e)]{Golenetskii2006e}Golenetskii, S., Aptekar, R., Mazets, E., et al., 2006e, GCN, 5748

\bibitem[Golenetskii et al.(2006f)]{Golenetskii2006f}Golenetskii, S., Aptekar, R., Mazets, E., et al., 2006f, GCN, 5837

\bibitem[Golenetskii et al.(2007a)]{Golenetskii2007a}Golenetskii, S., Aptekar, R., Mazets, E., et al., 2007a, GCN, 6403

\bibitem[Golenetskii et al.(2007b)]{Golenetskii2007b}Golenetskii, S., Aptekar, R., Mazets, E., et al., 2007b, GCN, 6849

\bibitem[Golenetskii et al.(2007c)]{Golenetskii2007c}Golenetskii, S., Aptekar, R., Mazets, E., et al., 2007c, GCN, 6879

\bibitem[Golenetskii et al.(2007d)]{Golenetskii2007d}Golenetskii, S., Aptekar, R., Mazets, E., et al., 2007d, GCN, 6960

\bibitem[Golenetskii et al.(2007e)]{Golenetskii2007e}Golenetskii, S., Aptekar, R., Mazets, E., et al., 2007e, GCN, 7114

\bibitem[Golenetskii et al.(2007f)]{Golenetskii2007f}Golenetskii, S., Aptekar, R., Mazets, E., et al., 2007f, GCN, 7155

\bibitem[Golenetskii et al.(2008a)]{Golenetskii2008a}Golenetskii, S., Aptekar, R., Mazets, E., et al., 2008a, GCN, 7487

\bibitem[Golenetskii et al.(2008b)]{Golenetskii2008b}Golenetskii, S., Aptekar, R., Mazets, E., et al., 2008b, GCN, 7589

\bibitem[Golenetskii et al.(2008c)]{Golenetskii2008c}Golenetskii, S., Aptekar, R., Mazets, E., et al., 2008c, GCN, 7812

\bibitem[Golenetskii et al.(2008d)]{Golenetskii2008d}Golenetskii, S., Aptekar, R., Mazets, E., et al., 2008d, GCN, 7854

\bibitem[Golenetskii et al.(2008e)]{Golenetskii2008e}Golenetskii, S., Aptekar, R., Mazets, E., et al., 2008d, GCN, 7862

\bibitem[Golenetskii et al.(2009)]{Golenetskii2009}Golenetskii, S., Aptekar, R., Mazets, E., et al., 2009, GCN, 10083

\bibitem[Golenetskii et al.(2010a)]{Golenetskii2010a}Golenetskii, S., Aptekar, R., Frederiks, D., et al., 2010a, GCN, 11021

\bibitem[Golenetskii et al.(2010b)]{Golenetskii2010b}Golenetskii, S., Aptekar, R., Frederiks, D., et al., 2010b, GCN, 11119

\bibitem[Golenetskii et al.(2010c)]{Golenetskii2010c}Golenetskii, S., Aptekar, R., Frederiks, D., et al., 2010c, GCN, 11251

\bibitem[Golenetskii et al.(2011a)]{Golenetskii2011a}Golenetskii, S., Aptekar, R., Mazets, E., et al., 2011a, GCN, 11971

\bibitem[Golenetskii et al.(2011b)]{Golenetskii2011b}Golenetskii, S., Aptekar, R., Frederiks, D., et al., 2011b, GCN, 12166

\bibitem[Golenetskii et al.(2011c)]{Golenetskii2011c}Golenetskii, S., Aptekar, R., Mazets, E., et al., 2011c, GCN, 12223

\bibitem[Golenetskii et al.(2011d)]{Golenetskii2011d}Golenetskii, S., Aptekar, R., Frederiks, D., et al., 2011d, GCN, 12270

\bibitem[Golenetskii et al.(2011e)]{Golenetskii2011e}Golenetskii, S., Aptekar, R., Frederiks, D., et al., 2011e, GCN, 12433

\bibitem[Golenetskii et al.(2013a)]{Golenetskii2013a}Golenetskii, S., Aptekar, R., Frederiks, D., et al., 2013a, GCN, 14368

\bibitem[Golenetskii et al.(2013b)]{Golenetskii2013b}Golenetskii, S., Aptekar, R., Frederiks, D., et al., 2013b, GCN, 14575

\bibitem[Golenetskii et al.(2013c)]{Golenetskii2013c}Golenetskii, S., Aptekar, R., Frederiks, D., et al., 2013c, GCN, 14958

\bibitem[Golenetskii et al.(2013d)]{Golenetskii2013d}Golenetskii, S., Aptekar, R., Frederiks, D., et al., 2013d, GCN, 15145

\bibitem[Golenetskii et al.(2013e)]{Golenetskii2013e}Golenetskii, S., Aptekar, R., Frederiks, D., et al., 2013e, GCN, 15203

\bibitem[Golenetskii et al.(2013f)]{Golenetskii2013f}Golenetskii, S., Aptekar, R., Frederiks, D., et al., 2013f, GCN, 15413

\bibitem[\protect\citeauthoryear{Greiner et al.}{2009}]{Greiner2009} Greiner, J., Kr\"{u}hler, T., Fynbo, J. P. U., et al., 2009, ApJ, 693, 1610

\bibitem[\protect\citeauthoryear{Greiner et al.}{2011}]{Greiner2011} Greiner, J., Kr\"{u}hler, T., Klose, S., 2011, A\&A, 526, 30

\bibitem[Gruber(2010)]{Gruber2010}Gruber, D., 2010, GCN, 11454

\bibitem[Gruber(2012a)]{Gruber2012a}Gruber, D., 2012a, GCN, 12874

\bibitem[Gruber(2012b)]{Gruber2012b}Gruber, D., 2012b, GCN, 13469

\bibitem[\protect\citeauthoryear{Guetta \& Piran}{2007}]{Guetta07}Guetta, D., \& Piran, T., 2007, J. Cosmol. Astropart. Phys., 7, 3

\bibitem[Guetta et al. (2005)]{Guetta05}Guetta, D., Piran, T., \& Waxman, E., 2005, ApJ, 619, 412

\bibitem[Hjorth et al. (2003)]{Hjorth2003}Hjorth, J., et al., 2003, Nature, 423, 847

\bibitem[Hjorth et al. (2012)]{Hjorth2012}Hjorth, J., Malesani, D., Jakobsson, P., et al., 2012, ApJ, 756, 187

\bibitem[Hopkins \& Beacon (2006)]{Hopkins2006}Hopkins, A. M., \& Beacon, J. F. 2006, ApJ, 651, 142

\bibitem[Howell et al. (2014)]{Howell14}Howell, E. J., Coward, D. M., Stratta, G., Gendre, B., \& Zhou, H., arXiv: 1407.2333

\bibitem[Kirshner et al. (1978)]{Kirshner1978}Kirshner, R. P., Oemler, A. \& Schechter, P. L., 1978, AJ, 83, 1549

\bibitem[Kistler et al. (2008)]{Kistler2008}Kistler, M. D., Y\"{u}ksel, H., Beacom, J. F. \& Stanek, K. Z., 2008, ApJ, 673, L119

\bibitem[Kistler et al. (2009)]{Kistler2009}Kistler, M. D., Y\"{u}ksel, H., Beacom, J. F., Hopkins, A. M. \& Wyithe, J. S. B., 2009, ApJ, 705, L104

\bibitem[Kocevski \& Liang (2006)]{Kocevski06}Kocevski, D., \& Liang, E., 2006, ApJ, 642, 371

\bibitem[\protect\citeauthoryear{Kouveliotou et al.}{1993}]{Kou93}Kouveliotou, C., Meegan, C., A., Fishman, G. J., et al., 1993, ApJ, 413, L101

\bibitem[Lamb \& Reichart (2000)]{LambReichart2000}Lamb, D. Q., \& Reichart, D. E. 2000, ApJ, 536, 1

\bibitem[Le \& Dermer (2007)]{Le07}Le, T., \& Dermer, C. D. 2007, ApJ, 661, 394

\bibitem[\protect\citeauthoryear{Li}{2008}]{Li08}Li, L. X. 2008, MNRAS, 388, 1487

\bibitem[Liang et al. (2007)]{Liang07}Liang, E. W., Zhang, B., Virgili, F., \& Dai, Z. G., 2007, ApJ, 662, 1111

\bibitem[Lilly et al. (1996)]{Lilly1996}Lilly, S. J., Lefevre, O., Hammer, F., \& Crampton, D. 1996, ApJ, 460, L1

\bibitem[Lin et al. (2004)]{Lin04}Lin, J. R., Zhang, S. N., \& Li, T. P., 2004, ApJ, 605, 819

\bibitem[Lloyd-Ronning et al. (2002)]{Lloyd-Ronning2002}Llyd-Ronning, N. M., Fryer, C. L. \& Ramirez-Ruiz, E., 2002, ApJ, 574, 554

\bibitem[Loh \& Spillar (1986)]{Loh1986}Loh, E. D. \& Spillar, E. J., 1986, ApJ, 307, L1

\bibitem[L\"{u} et al.(2014)]{Lv14}L\"{u}, H., Zhang, B., Liang, E. W., Zhang, B. B., \& Sakamoto,
T., 2014, MNRAS, 442, 1922

\bibitem[Lynden-Bell(1971)]{Lynden-Bell1971}Lynden-Bell, D., 1971, MNRAS, 155, 95

\bibitem[Maloney \& Petrosian (1999)]{MaloneyPetrosian1999}Maloney, A., \& Petrosian, V. 1999, ApJ, 518, 32

\bibitem[Mazzali et al.(2006)]{Mazzali06}Mazzali, P. A., Deng, J., Nomoto, K., et al. 2006, Nature, 442, 1018

\bibitem[McBreen(2009)]{McBreen2009}McBreen, S., 2009, GCN, 9415

\bibitem[McGlynn(2012)]{McGlynn2012}McGlynn, S., 2012, GCN, 14012

\bibitem[\protect\citeauthoryear{McQuinn et al.}{2008}]{McQuinn08}McQuinn, M., Lidz, A., Zaldarriaga, M., Hernquist, L. \& Dutta, S., 2008, MNRAS, 388, 1101

\bibitem[Merighi et al.(1986)]{Merighi1986}Merighi, R., Marano, B. \& Vettolani, G., 1986, A\&A, 160, 398

\bibitem[\protect\citeauthoryear{M\'{e}sz\'{a}ros}{2006}]{Meszaros06} M\'{e}sz\'{a}ros, P., 2006, Rep. Prog. Phys., 69, 2259

\bibitem[Nava et al.(2011)]{Nava2011}Nava, L., Ghirlanda, G., Ghisellini G., et al., 2011, A\&A, 530, A21

\bibitem[Nava et al.(2012)]{Nava2012}Nava, L., Salvaterra, R., Ghirlanda, G., et al., 2012, MNRAS, 421, 1256

\bibitem[Ohno et al.(2007)]{Ohno2007}Ohno, M., Uehara, T., Takahashi, T., et al., 2007, GCN, 6638

\bibitem[Ohno et al.(2008)]{Ohno2008}Ohno, M., Kokubun, M., Suzuki, M., et al., 2008, GCN, 7630

\bibitem[Pal'Shin et al.(2008)]{Pal'Shin2008}Pal'Shin, V., Golenetskii, S., Aptekar, R., Mazets, E., et al., 2008, GCN, 8256

\bibitem[Pal'Shin et al.(2013)]{Pal'Shin2013}Pal'Shin, V., Golenetskii, S., Aptekar, R., Mazets, E., et al., 2013, GCN, 14702

\bibitem[P\'{e}langeon et al.(2008)]{Pelangeon08}P\'{e}langeon, A., et al., 2008, A\&A, 491, 157.

\bibitem[Pelassa(2011)]{Pelassa2011}Pelassa, V., 2011, GCN, 12545

\bibitem[Perley et al.(2013)]{Perley2013}Perley, D. A., Levan, A. J., Tanvir, N. R., et al., 2013, ApJ, 778, 128

\bibitem[Peterson et al.(1986)]{Peterson1986}Peterson, B. A., Ellis, R. S., Efstathiou, G., Shanks, T., Bean, A. J., Fong, R. \& Zen-Long, Z., 1986, MNRAS, 221, 233

\bibitem[Petrosian (1993)]{Petrosian1993}Petrosian, V. 1993, ApJ, 402, L33

\bibitem[Petrosian et al. (2009)]{Petrosian2009}Petrosian, V., Bouvier, A. \& Ryde, F., 2009, arXiv:0909.5051v1

\bibitem[Pian et al. (2006)]{Pian06}Pian, E., Mazzali, P. A., Masetti, N., et al. 2006, Nature, 442, 1011

\bibitem[Porciani \& Madau (2001)]{PorcianiMadau2001}Porciani, C., \& Madau, P. 2001, ApJ, 548, 522

\bibitem[Qin et al.(2010)]{Qin10}Qin, S. F., Liang, E. W., Lu, R. J., Wei, J. Y., \& Zhang, S. N. 2010, MNRAS, 406, 558

\bibitem[Sakamoto et al.(2006)]{Sakamoto2006}Sakamoto, T., Barbier, L., Barthelmy, S., et al., 2006, GCN, 5029

\bibitem[Sakamoto et al.(2008)]{Sakamoto2008}Sakamoto, T., Barthelmy, S. D., Baumgartner, W., et al., 2008, GCN, 8101

\bibitem[Sakamoto et al.(2011)]{Sakamoto2011}Sakamoto, T., Barthelmy, S. D., Baumgartner, W., et al., 2011, GCN, 12276

\bibitem[\protect\citeauthoryear{Salvaterra \& Chincarini}{2007}]{Salvaterra07} Salvaterra, R., \& Chincarini, G., 2007a, ApJ, 656, L49

\bibitem[Salvaterra et al. (2009a)]{Salvaterra09}Salvaterra, R., Della Valle, M., Campana, S., et al., 2009, Nature, 461, 1258

\bibitem[Salvaterra et al. (2009b)]{Salvaterra09b}Salvaterra, R., Guidorzi, C., Campana, S., Chincarini, G., \& Tagliaferri, G., 2009, MNRAS, 396, 299

\bibitem[Salvaterra et al. (2012)]{Salvaterra2012}Salvaterra, R., Campana, S., Vergani, S. D., et al., 2012, ApJ, 749, 68

\bibitem[\protect\citeauthoryear{Schaefer}{2007}]{Schaefer07} Schaefer B.~E., 2007, ApJ, 660, 16

\bibitem[Schmidt(1999)]{Schmidt99}Schmidt, M., 1999, ApJ, 523, L117

\bibitem[Soderberg et al.(2006)]{Soderberg06}Soderberg, A. M., Kulkarni, S. R., Nakar, E., et al. 2006, Nature, 442, 1014

\bibitem[Stamatikos et al.(2006)]{Stamatikos2006}Stamatikos, M., Barbier, L., Barthelmy, S., et al., 2006, GCN, 5289

\bibitem[Stamatikos et al.(2008)]{Stamatikos2008}Stamatikos, M., Barthelmy, S. D., Cummings, J., et al., 2008, GCN, 7277

\bibitem[Stanek et al.(2003)]{Stanek2003}Stanek, K. Z., Matheson, T., Garnavich, P. M., et al., 2003, ApJ, 591, L17

\bibitem[Sugita et al.(2009)]{Sugita2009}Sugita, S., Yamaoka, K., Ohno, M., et al., 2009, PASJ, 61, 521

\bibitem[Tan et al.(2013)]{Tan14}Tan, W. W., Cao, X. F., \& Yu, Y. W., 2013, ApJL, 772, L8

\bibitem[Tanvir et al.(2009)]{Tanvir09}Tanvir, N. R., Fox, D. B., LevanA, J., et al., 2009, Nature, 461, 1254

\bibitem[Totani(1997)]{Totani97}Totani, T., 1997, ApJ, 486, L71

\bibitem[\protect\citeauthoryear{Totani et al.}{2006}]{Totani06}Totani, T., Kawai, N., Kosugi, G., et al. 2006, PASJ, 58, 485

\bibitem[Tueller et al.(2008)]{Tueller2008}Tueller, J., Barthelmy, S. D., Baumgartner, W., et al., 2008, GCN, 7604

\bibitem[van der Horst(2010)]{Horst2010}van der Horst, A. J., 2010, GCN, 11477

\bibitem[\protect\citeauthoryear{Vergani et al.}{2014}]{Vergani2014}Vergani, S. D., Salvaterra, R., Japelj, J., et al., 2014, arXive:1409.7064

\bibitem[von Kienlin(2013)]{Kienlin2013}von Kienlin, A., 2013, GCN, 14473

\bibitem[\protect\citeauthoryear{Virgili et al.}{2011}]{Virgili11}Virgili, F. J., Zhang, B., Nagamine, K., \& Choi, J. H. 2011, MNRAS, 417, 3025

\bibitem[Wanderman \& Piran (2010)]{Wanderman2010}Wanderman, D. \& Piran, T. 2010, MNRAS, 406, 1944

\bibitem[\protect\citeauthoryear{Wang et al.}{2012}]{Wang12} Wang, F. Y., Bromm, V., Greif. T. H., Stacy, A., Dai, Z. G., Loeb, A. \& Cheng, K. S., 2012, ApJ, 760, 27

\bibitem[\protect\citeauthoryear{Wang \& Dai}{2009}]{Wang09} Wang, F. Y., \& Dai, Z. G. 2009, MNRAS, 400, L10

\bibitem[\protect\citeauthoryear{Wang \& Dai}{2011}]{WangD11} Wang, F. Y., \& Dai, Z. G. 2011a, ApJL, 727, 34

\bibitem[\protect\citeauthoryear{Wang \& Dai}{2014}]{WangD14} Wang, F. Y., \& Dai, Z. G. 2014, ApJS, 213, 15

\bibitem[Wang et al. (2014)]{Wang14}Wang, F. Y., Dai, Z. G., \& Liang, E. W.,
2015, arXiv: 1504.00735

\bibitem[Wang et al. (2011)]{Wang11}Wang, F. Y., Qi, S. \& Dai, Z. G. 2011, MNRAS, 415, 3423

\bibitem[Wang (2013)]{Wang13}Wang, F. Y., 2013, A\&A, 556, A90

\bibitem[Wijers et al. (1998)]{Wijers1998}Wijers, R. A. M., Bloom, J. S., Bagla, J. S. \& Natarajan, P., 1998, MNRAS, 294, L13

\bibitem[\protect\citeauthoryear{Woosley}{1993}]{Woosley93}Woosley, S. E. 1993, ApJ, 405, 273

\bibitem[Wu et al.(2012)]{Wu2012}Wu, S. W., Xu, D., Zhang, F. W. \& Wei, D. M., 2012, MNRAS, 423, 2627

\bibitem[Xiong(2011)]{Xiong2011}Xiong, S., 2011, GCN, 12287

\bibitem[Xiong \& Rau(2013)]{Xiong2013}Xiong, S. \& Rau, A., 2013, GCN, 14429

\bibitem[Yonetoku et al.(2004)]{Yonetoku2004}Yonetoku, D., Murakami, T., Nakamura, T., Yamazaki, R., Inoue, A. K. \& Ioka, K., 2004, ApJ, 609, 935

\bibitem[Yonetoku et al.(2014)]{Yonetoku14}Yonetoku, D., Nakamura, T., Sawano, T., Takahashi, K., Toyanago, A., 2014, ApJ, 789, 65

\bibitem[Younes \& Barthelmy(2012)]{Younes2012}Younes, G. \& Barthelmy, S. D., 2012, GCN, 13722

\bibitem[Younes(2012)]{Younes2012a}Younes, G., 2012, GCN, 13809

\bibitem[Younes \& Bhat(2013)]{Younes2013}Younes, G. \& Bhat, P. N., 2013, GCN, 14219

\bibitem[Yu(2012)]{Yu2012}Yu, D., 2012, GCN, 14078

\bibitem[Y\"{u}ksel et al. (2008)]{Yuksel2008}Y\"{u}ksel, H., Kistler, M. D., Beacom, J. F. \& Hopkins, A. M., 2008, ApJ, 683, L5

\bibitem[\protect\citeauthoryear{Zhang}{2006}]{Zhang06} Zhang, B., 2006, Nature, 444, 1010

\bibitem[\protect\citeauthoryear{Zhang}{2007}]{Zhang07} Zhang, B., 2007, Chin.~J.~Astron.~Astrophys., 7, 1

\bibitem[\protect\citeauthoryear{Zhang et al.}{2007}]{ZhangB07}Zhang, B., Liang, E. W., Page, K. L., et al. 2007, ApJ, 655, 989

\bibitem[\protect\citeauthoryear{Zhang et al.}{2009}]{Zhang09}Zhang, B., Zhang, B. B., Virgili, F. J., et al., 2009, ApJ, 703, 1696


\end{thebibliography}
\end{document}